\documentclass[12pt]{article}
\usepackage{graphicx}
\usepackage{amsmath}
\usepackage{amsfonts}
\usepackage{amssymb}
\usepackage{bm}

\usepackage[
  backend=biber,
  bibencoding=utf8,     
  style=numeric-comp,  
  sorting=none,        
  sortcites=true,      
  maxbibnames=99
]{biblatex}
\renewbibmacro{in:}{}

\usepackage[english]{babel}
\usepackage[latin1]{inputenc}
\usepackage{times}
\usepackage[T1]{fontenc}

\newcommand{\beq}{\begin{equation}}
\newcommand{\eeq}{\end{equation}}
\newcommand{\beqa}{\begin{eqnarray}}
\newcommand{\eeqa}{\end{eqnarray}}

\newcommand{\RR}{\ensuremath{ \mathbb{R}} }

\newcommand{\NN}{\ensuremath{ \mathbb{N}} }

\newcommand{\D}{{\rm d}}

\newcommand{\Lcal}{\mathcal{L}}

\def\beqa*{\begin{eqnarray*}}
\def\eeqa*{\end{eqnarray*}}

\newcommand{\realp}{{\mathbb{R}}{\rm e}}

\usepackage{tikz}
\usetikzlibrary{arrows}
\tikzstyle{block}=[draw opacity=0.7,line width=1.4cm]

\newtheorem{theorem}{Theorem}

\newcounter{llista}

\begin{document}
\title{Nonlocal Mechanics}
\author{
Carlos Heredia\thanks{e-mail address: carlosherediapimienta@gmail.com} \\
\textit{IAMM Research, Department of Applied Artificial Intelligence} \\
\textit{DAMM, Carrer del Rossell\'o 515, 08025 Barcelona, Catalonia, Spain}
\and
Josep Llosa\thanks{e-mail address: pitu.llosa@ub.edu} \\
\textit{Faculty of Physics (FQA and ICC)} \\
\textit{University of Barcelona, Diagonal 645, 08028 Barcelona, Catalonia, Spain}
}
\maketitle

\begin{abstract}
 We introduce a Hamiltonian framework for nonlocal Lagrangian systems without relying on infinite-derivative expansions. Starting from a (trajectory-based) variational principle and a generalized Noether theorem, we define the canonical momenta and energy. Moreover, we construct a (pre)symplectic form on the kinematic space, and show that its restriction to the phase space (by implementing the constraints) yields a true (pre)symplectic structure encoding the dynamics. Three examples ---a finite nonlocal oscillator, the fully nonlocal Pais-Uhlenbeck model, and a delayed harmonic oscillator--- demonstrate how phase space and the Hamiltonian emerge without explicitly solving the Euler-Lagrange equations.
\end{abstract}

\section{Introduction}\label{S1}
The canonical formalism for nonlocal systems has long been regarded as conceptually delicate and technically cumbersome.  In an extended series of articles\,\cite{Llosa1987,Vives1994,Gomis2002,Heredia2021a,Heredia2022}, we have progressively clarified these issues for nonlocal Lagrangians and their subsequent quantization.  The present paper synthesizes those developments and, crucially, incorporates several advances achieved during its preparation. The most important of these is a novel, fully mechanical recipe that extracts the canonical phase-space coordinates from the Hamiltonian formalism directly, dispensing with any prior imposition of dynamical solutions ---an assumption that had been a weak point in earlier versions.  In one fell swoop, the method identifies the degrees of freedom of a nonlocal theory and equips them with the appropriate (pre)symplectic structure.

This new viewpoint immediately reframes the role of the Euler-Lagrange (EL) equations. Whereas our earlier constructions treated the EL equations as \emph{ad hoc} primary constraints, we now show that they arise \emph{dynamically} as secondary constraints dictated by Hamilton's equations. The result closes the logical gap between the variational and Hamiltonian descriptions and paves a direct path to quantization.

Throughout the exposition we highlight the close parallel between the standard local theory and its nonlocal extension: variational principle, boundary terms, Legendre map, Hamiltonian formalism and, finally, the equivalence between EL and Hamilton's equations.  This deliberate mirroring underscores that the framework proposed here is a natural ---albeit non-trivial--- continuation of standard techniques.

Sections~\ref{S2} and~\ref{S3} review the canonical treatment of first-order and higher-order ($n>1$) \emph{local} Lagrangians, respectively. Our approach differs, however, from the classical treatment found in standard text books, e.g.~\cite{Goldstein,Gantmacher,Landau}, where the Lagrangian is required to be regular ---equivalently, the Legendre transform must be invertible--- before introducing the Hamiltonian formalism. By contrast, we exploit presymplectic differential 2-form \cite{Godbillon,Marsden} to push the analysis as far as possible without assuming the full independence of the canonical momenta, i.e., without first removing constraints. This yields an equation that links the Hamiltonian, a presymplectic 2-form, and the Hamiltonian vector field that generates time evolution ---an immediate precursor to Hamilton's equations. When the Lagrangian is regular, the 2-form becomes non-degenerate and hence symplectic (the dual of the Poisson bracket), and our precursor equation collapses to the familiar Hamilton equations.

Section~\ref{S4} analyses \emph{infinite-order} Lagrangians.  Historically, the latter were invoked as surrogates for nonlocal models ---one simply replaced the finite order $n$ by $\infty$ in every formula.  Such an approach appears already in the early works on direct interaction electrodynamics\,\cite{Tetrode1922,Fokker,WheelerFeynman}, in the Pais--Uhlenbeck oscillator\,\cite{PaisUhlenbeck1950}, in the electrodynamics of dispersive media \cite{JLlosaMT}, and more recently in string field theory, non-commutative field theory, fractional-derivative Maxwell theory, and nonlocal gravity\,\cite{Yukawa,Moller,EliezerWoodard1989,Witten1986,Biswas2012,HerediaKolar2022,BoosKolar2021,CalcagniModesto2014,CapozzielloBajardi2022,Mashhoon2017,Heydeman2023} to tame ultraviolet divergences.  We devote a full section to these Lagrangians both  because of their widespread use and because their naive use masks serious mathematical inconsistencies\footnote{In fact, our initial approach was to employ infinite series ---as is common in the literature; however, we were mathematically criticized for that rather naive use of infinite series. This inspired us to adopt a summability approach for these series, as detailed in Ref.~\cite{JLlosaMT-ArXiv}.}.  In particular: (a) their EL equations are \emph{infinite-order} ordinary differential equations, raising subtle questions about initial data and the definition of phase space; (b) the Legendre transformation fails ---one cannot replace ``half of the derivatives'' by canonical momenta when ``half of infinity'' is still infinity.

The \emph{nonlocal} case is addressed in Section~\ref{S5}.  Here the Lagrangian depends functionally on the full or bounded-segment trajectory, without ever re-expressing it as an infinite-order derivative series. Instead, we work directly with functional techniques, which keep both the conceptual picture transparent and the calculations manageable. We first reformulate the variational principle, arguing that the action must extend over the entire real line for consistency (Section~\ref{Paradox}).  Because integration by parts is no longer available, the canonical momenta must be defined anew.  Our strategy is to generalise Noether's theorem to nonlocal functionals, read the conserved charges, and thereby \emph{infer} an appropriate definition of the canonical momenta for nonlocal theories.

Having established the energy, the presymplectic form and the generator of time evolution, we derive a (pre-)Hamiltonian condition analogous to that in Sections~\ref{S2}-\ref{S3}. The remaining task is to exhibit suitable canonical coordinates and to identify the phase space. Sections~\ref{sec:nonlocal-oscillator}-\ref{Exemple2} illustrate the procedure with several explicit examples, demonstrating that the EL equations appear naturally as secondary constraints to define the constraint submanifold.

Finally, although our point of departure is firmly rooted in mathematical and theoretical physics, the exposition is organised as a self-contained handbook. Every conceptual ingredient is accompanied by concrete prescriptions so that researchers in \emph{any} discipline can deploy the nonlocal extension of their theory in a systematic way. In this spirit the present work aspires to be not merely a specialised treatise, but a practical toolkit that places the power of nonlocal Lagrangian, Hamiltonian and symmetry methods within immediate reach of the broader scientific community.

\bigskip
\noindent\textbf{Notation.}  In an expression containing some variables, the $=$ sign is a polysemic one. It may be used as: (a) an equation, that only holds when the variables are replaced by the solutions, (b) a definition, where the right hand side defines the symbol in left hand side, or (c) an identity, when both sides are equal no matter what values are given to the variables. 
In the present text, when necessary to avoid confusion, sometimes we use different symbols for each of these instances and write $:=$ to indicate ``definition'' or $\equiv$ to mean ``identically equal''. The symbol $=$ is then reserved to equations. However, to avoid pedantry as much as possible, this will not be the rule. 

\section{Standard Lagrangian systems. An outline \label{S2}}
This section briefly summarizes the key concepts of standard local Lagrangian mechanics. Consider a system characterized by a Lagrangian function $L(q, \dot{q}, t)$ with $m$ degrees of freedom. The associated action integral is defined as the functional
\beq \label{e1}
S([q],t_1,t_2) := \int_{t_1}^{t_2} L\left(q(t), \dot q(t), t\right)\,\D t  
\eeq
acting upon \emph{kinematic trajectories}, specifically $q \in \mathcal{C}^2\left([t_1, t_2], \mathcal{U}\right)$, with $\mathcal{U} \subset \mathbb{R}^m$ an open set representing the configuration space. The notation $[q]$ emphasizes functional dependence, and the point $q = (q^\alpha)_{\alpha = 1, \dots, m}$; however, for clarity and simplicity, we shall generally omit superindices unless explicitly required.

The \emph{dynamical trajectory} connecting points $(q_1, t_1)$ and $(q_2, t_2)$ is defined as the trajectory that makes the action integral stationary while satisfying the boundary conditions $q(t_1) = q_1$ and $q(t_2) = q_2$. Formally, it satisfies the variational principle:
\begin{list}
{{\bf (\alph{llista})}}{\usecounter{llista}}
\item $\quad \delta S([q],t_1,t_2) = 0 \,, \qquad $for those variations $\delta q(t)$ such that
\item $\quad  \delta q(t_1)=  \delta q(t_2)= 0 \,.$ 
\end{list}
Explicitly, this condition reads:
\beq \label{e2}
 \delta S \equiv \int_{t_1}^{t_2} \left[\frac{\partial L}{\partial q}\, \delta q(t) + \frac{\partial L}{\partial \dot q}\, \delta \dot q(t) \right]\,\D t = 0 \,, 
\eeq
where summation over repeated indices is implied in expressions such as $\displaystyle{\frac{\partial L}{\partial q} \delta q(t) = \frac{\partial L}{\partial q^j} \delta q^j(t)}$.

As a consequence of condition \textbf{(b)}, $\delta \dot{q}(t)$ cannot be arbitrary, requiring integration by parts to rewrite expression (\ref{e2}) as:
\beq \label{e3}
 \delta S \equiv \int_{t_1}^{t_2} \D t\, \delta q(t)\,\mathcal{E}\left(q(t), \dot q(t), \ddot q(t), t\right) + \left[ p(t)\,\delta q(t) \right]_{t_1}^{t_2} = 0 \,,
\eeq
where
\begin{eqnarray} \label{e4a}
 & &\mathcal{E}\left(q,\dot q, \ddot q, t\right) := \frac{\partial L}{\partial q} - \mathbf{D} p(q,\dot q, t) \,, \qquad \qquad p(q,\dot q, t) := \frac{\partial L \left(q,\dot q,t \right)}{\partial \dot q} \,,
\\[2ex]   \label{e4b}
 {\rm and}  & &  \mathbf{D}:= \dot q\,\frac{\partial \;}{\partial q} + \ddot q\,\frac{\partial \;}{\partial \dot q} + \frac{\partial \;}{\partial t} 
\end{eqnarray}
represents the infinitesimal generator of time evolution, also known as the total time derivative. 

Conditions \textbf{(a)} and \textbf{(b)}, together with the variational equation (\ref{e3}), imply the EL equations:
$$\mathcal{E}\left(q(t), \dot q(t), \ddot q(t), t\right) = 0\,, \qquad q(t_a) = q_a \,, \quad a=1,2\,.$$ 
For regular Lagrangians, satisfying $\,\displaystyle{\det\left( \frac{\partial^2 L }{\partial \dot q^\nu \dot q^\alpha} \right) \neq 0} \,$, the EL equations can be explicitly solved for accelerations, yielding an ordinary differential system in normal form:
\begin{equation}  \label{e4c} 
\ddot q^\nu = Q^\nu\left(q, \dot q, t\right) \,. 
\end{equation}
Provided the functions $Q^\nu$ satisfy appropriate mild conditions within a domain, standard existence and uniqueness theorems \cite{Barbu2016, Arnold1992} guarantee a unique solution in a neighborhood of any given ``initial" point $(q_1,\dot q_1, t_1)$ belonging in that domain. Thus, a unique solution to (\ref{e4c}) exists  
\begin{equation}  \label{e4d} 
\,q^\nu(t) = \varphi^\nu(t-t_1;q_1,\dot q_1, t_1)\,, \quad |t-t_1| < T\,,\quad \mbox{such that} \;\, q(t_1) = q_1\,, \;\,\text{and}  \;\, \dot q(t_1)= \dot q_1  \,.  
\end{equation}

Another important consequence of these theorems is the translational equality $q(\tau+t_1)=\varphi^\nu(\tau;q_1,\dot q_1, t_1)\,,$ which will play a central role in subsequent sections. This equality establishes a one-to-one correspondence between initial data $(q_1,\dot q_1,t_1)$ and dynamical trajectories. Thus, the set of dynamical trajectories is a $(2m+1)$-dimensional submanifold embedded within the space of kinematic trajectories $\mathcal{K}$ (with $m$ the number of degrees of freedom). Moreover, it is homeomorphic to the extended initial-data space $\mathcal{D}'=T\mathcal{U}\times\mathbb{R}$.

Finally, it is a well-known and easily proven result that the right-hand side of the EL equation identically vanishes if, and only if, the Lagrangian is a total derivative, i.e.\
\[
L(q,\dot q,t)=\mathbf{D}W(q,t).
\]
Furthermore, because the Euler-Lagrange operator $\mathcal{E}$ depends linearly on the Lagrangian, two Lagrangians $L_1$ and $L_2$ yield identical EL equations if and only if their difference is a total derivative:
\begin{equation}\label{e5}
L_1-L_2=\frac{\mathrm{d}W}{\mathrm{d}t},\qquad W=W(q,t) \,.
\end{equation}

\subsection{Noether Theorem  \label{S2.1}}
Consider the infinitesimal transformation:
\begin{equation}  \label{e6}
t^\prime(t) = t + \delta T(t) \,, \qquad \quad q^{\prime}(t^\prime) = q(t) + \delta q(t) \,,
 \end{equation}
which transforms an arbitrary time interval $[t_1,t_2]$ into $[t^\prime_1,t^\prime_2]$. The transformation of the velocities is obtained by differentiating the second equation (\ref{e6}), yielding:
\begin{equation}  \label{e6a}
\dot q^\prime (t^\prime)= \dot q(t) \,\left[1 - \delta\dot T(t) \right] + \delta \dot q(t) \,.
\end{equation} 
All equalities involving infinitesimal quantities are meant up to first-order terms, i.e. second-order and higher-order infinitesimals are neglected.
 
The induced transformation on the Lagrangian is defined by imposing invariance of the action integral for every kinematic trajectory. Explicitly, this invariance is expressed as:
\begin{equation}  \label{e7a}
\int_{t_1}^{t_2} \D t\, L\left(q(t), \dot q(t),t \right) \equiv \int_{t^\prime_1}^{t^\prime_2} \D t^\prime\,
L^\prime\left(q^\prime(t^\prime),\dot q^\prime (t^\prime), t^\prime\right)   \,,
\end{equation}
which directly implies:
\begin{equation}  \label{e7}
L^\prime\left(q^\prime(t^\prime),\dot q^\prime (t^\prime), t^\prime\right) = \frac{\D t}{\D t^\prime}\,L\left(q(t), \dot q(t),t \right) \,.
\end{equation}  

If the Lagrangian remains unchanged under the transformation (\ref{e6}-\ref{e7}), specifically satisfying
$$ L^\prime (q^\prime,\dot q^\prime,t^\prime) =  L(q^\prime,\dot q^\prime,t^\prime)  \,, $$
the transformation is termed a \emph{symmetry transformation}. More generally, a transformation preserving the EL equations without necessarily leaving the Lagrangian invariant is called a \emph{Noether symmetry}. In this case, the transformation modifies the Lagrangian by adding a total time derivative, as noted previously:
\begin{equation}  \label{e7b}
L^\prime\left(q^\prime,\dot q^\prime, t^\prime\right) = L \left(q^\prime,\dot q^\prime, t^\prime\right)+ \frac{\D W\left(q^\prime, t^\prime\right)}{\D t^\prime}  \,.
\end{equation} 

In the case of a Noether symmetry, combining equations (\ref{e7}) and (\ref{e7b}), the invariance of the action integral (\ref{e7a}) implies that, for any $t_1$ and $t_2\,$,
$$ \int_{t_1}^{t_2} \left[ \D t\,\left\{ L\left(q^\prime(t), \dot q^\prime(t),t \right) -  L\left(q(t), \dot q(t),t \right) \right\}  + \D W\left(q(t), t\right)  \right] + L_2\,\delta T_2 - L_1\,\delta T_1 \equiv 0 \,,$$ 
where $L_b := L(q_b,\dot q_b, t_b)$ and $\delta T_b := \delta T(t_b)\,.$ This expression can be reorganized as:
\beq  \label{e8}
\int_{t_1}^{t_2} \D t\,\left\{ L\left(q^\prime(t), \dot q^\prime(t),t) \right) -  L\left(q(t), \dot q(t),t \right)+ \frac{\D }{\D t} \left[ W\left(q(t), t\right) + L \,\delta T \right]  \right\}  \equiv 0 \,.
\eeq
Since equation (\ref{e8}) must hold for arbitrary $t_1$ and $t_2\,$, the integrand must vanish identically, leading to:
$$ L\left(q^\prime(t), \dot q^\prime(t),t \right) -  L\left(q(t), \dot q(t),t \right) + \frac{\D }{\D t} \left[ W\left(q(t), t\right) + L \,\delta T \right]    \equiv 0 \,. $$
Including the expressions (\ref{e6}) and (\ref{e6a}) and using the shorthand notation$\, \tilde\delta q := q^\prime(t) - q(t) = \delta q - \dot q \delta T \,$, this condition becomes
$$ \frac{\partial L}{\partial q}\,\tilde\delta q + \frac{\partial L}{\partial \dot q}\,\frac{\D\,\tilde\delta q}{\D t} + \frac{\D (L\,\delta T + W)}{\D t} \equiv 0 \,,$$
which leads directly to the \emph{Noether identity}:
\beq  \label{e8a}
 \mathcal{E}(q)\,\tilde\delta q  + \frac{\D }{\D t} \left(p\,\delta q - E\,\delta T + W \right) \equiv 0 \,,
\eeq
where 
\beq  \label{e8b}
p := \frac{\partial L}{\partial \dot q} \qquad {\rm and} \qquad E := p\,\dot q - L \,,
\eeq
denote the canonical momentum and energy, respectively.

As a direct corollary of the Noether identity (\ref{e8a}), it follows that the quantity
\beq  \label{e8c}
Q:= p\,\delta q - E\,\delta T + W 
\eeq
is conserved along the dynamical trajectories. That is, whenever the EL equations $\mathcal{E}(q) = 0\,$ are satisfied,  the quantity $Q$ remains constant in time, and thus constitutes a conserved quantity or \emph{integral of motion}.

\subsection{Hamiltonian formalism \label{S2.2}}
Our approach to the subject departs slightly from standard textbook treatments~\cite{Goldstein,Gantmacher,Landau}. In expression (\ref{e8c}), the roles of the canonical momentum and the energy~\eqref{e8b} in defining the conserved quantity $Q$ are made explicit: they are
\beq  \label{e9}
   p(q,\dot q, t) := \frac{\partial L(q,\dot q, t)}{\partial \dot q}  \qquad {\rm and} \qquad E(q,\dot q,t) := p(q,\dot q, t)\dot q - L(q,\dot q, t)  \, , 
\eeq
and appear as the prefactors of the infinitesimal variations $\delta q$ and  $\delta T\,$. These quantities correspond to the canonical momentum and the energy of the system, respectively.

In contrast to standard treatments, we do not require the Hessian $\displaystyle{\frac{\partial^2 L}{\partial \dot q_\alpha \partial \dot q_\beta}}\,$ to be regular, nor do we demand that equation (\ref{e9}) be solvable for the velocities. Instead, our focus shifts to the differential 2-form 
\beq  \label{e9b}
  \beta := \omega - \D E \wedge \D t \,,\qquad {\rm with} \qquad \omega := \D p(q,\dot q,t)\wedge \D q \,,   
\eeq
that are defined on the extended initial data space, $\,\mathcal{D}^\prime := \mathcal{D}\times \RR\,$. The 2-form $\,\omega\,$ is closed and is known as the \emph{presymplectic form}; the prefix emphasizes that its rank might not be maximal. This terminology is natural and justified, particularly to readers familiar with symplectic geometry and classical mechanics~\cite{Godbillon,Marsden,Flanders}.

The inner product of the time evolution generator $\, \mathbf{D}\,$ ---as defined in equation (\ref{e4b})--- with the differential form $\beta$ is computed as follows:
\begin{eqnarray*}
i_\mathbf{D} \beta & \equiv & \mathbf{D} p\,\D q - \dot q\,\D p - \mathbf{D} E \,\D t + \D E  \\[1.5ex]
   & \equiv & \mathbf{D}p\,\D q - \dot q\,\D p - \mathbf{D}E \,\D t + p \,\D \dot q + \dot q\,\D p - \partial_t L - \frac{\partial L}{\partial q}\,\D q - \frac{\partial L}{\partial \dot q}\,\D\dot q \\[1.5ex] 
   & \equiv & \left(\mathbf{D}p -\frac{\partial L}{\partial q} \right)\, \D q - \left(\mathbf{D} E +\frac{\partial L}{\partial t} \right)\, \D t \,
\end{eqnarray*}
and, using that
$$  \mathbf{D} E \equiv \mathbf{D}p\,\dot q + p\,\ddot q - \dot q\,\frac{\partial L}{\partial q} - \ddot q \,\frac{\partial L}{\partial \dot q} - \frac{\partial L}{\partial t} \equiv  
\left(\mathbf{D}p -\frac{\partial L}{\partial q} \right)\, \dot q - \frac{\partial L}{\partial t}  \,,
$$
we easily arrive at
$$ i_\mathbf{D} \beta \equiv - \mathcal{E} \,\left(\D q - \dot q\,\D t \right) \,,$$
where $\,\mathcal{E}=0\,$ is the EL equation (\ref{e4a}). Therefore,
\beq  \label{e9c}
   i_\mathbf{D} \beta = 0 \qquad \quad \Leftrightarrow \qquad \quad \mathcal{E} = 0 \,, 
\eeq
that is, the EL equation is equivalent to the condition
\beq  \label{e9d}
  i_\mathbf{D} \beta = 0 \,.
\eeq
This establishes a geometric characterization of the dynamics through the vanishing of the contraction of $\beta$ with the generator $\mathbf{D}$.

So far, we have not required the Lagrangian to be regular. If regularity holds, equation (\ref{e9}) defines a Legendre transformation $(q,\dot q,t)\leftrightarrow(q,p,t)$, which can be inverted to express velocities explicitly in terms of the canonical variables as $\dot q = \dot q(q,p,t)$. Under these conditions, the extended phase space can be coordinatized by the independent canonical variables $(q, p, t)\,$. With regularity, the 2-form $\omega=\mathrm{d}p\wedge\mathrm{d}q$ has maximal rank $2m$, and the 2-form $\beta$ defines a contact structure~\cite{Godbillon} on the extended phase space $T^\ast\mathcal{U}\times\mathbb{R}$. Consequently, equation (\ref{e9d}) admits a unique solution for the time-evolution generator $\,\mathbf{D}\,$.

The Hamiltonian is defined as the energy written in terms of the canonical variables:
$$\,H(q,p,t) := E (q, \dot q(q,p,t),t)\,,$$
 where $\dot q(q,p,t)$  is obtained from the inverse Legendre transformation. Substituting this into equation (\ref{e9d}) yields
$$ i_\mathbf{D}\omega + \D H + \partial_t L\,\D t = 0 \,, $$ 
which amounts to the Hamilton equations:
\beq  \label{e9e}
\frac{\D q}{\D t} = \frac{\partial H}{\partial p} \,, \qquad \frac{\D p}{\D t} = - \frac{\partial H}{\partial q}  \,, \qquad \mbox{and the well known relation} \qquad 
\frac{\partial H}{\partial t} = - \frac{\partial L}{\partial t} \,.
\eeq

\section{Higher-order Lagrangian systems. An outline \label{S3}}
If the Lagrangian
\[
L\bigl(q^{(j)},t\bigr)\;=\;L\bigl(q,\dot q,\ldots,q^{(n)},t\bigr),
\]
depends on the time-derivatives of the configuration variables only up to a finite order \(n\), the development parallels the familiar first-order case \((n=1)\). We denote by \(q^{(\ell)}\) the \(\ell\)-th time derivative of \(q\) for \(\ell=0,\ldots,n\). The action integral is the functional
\beq \label{en1}
S([q],t_1,t_2) := \int_{t_1}^{t_2} L\left(q(t),q^{(1)}(t) \ldots q^{(n)}(t), t\right)\,\D t  
\eeq
and is defined on {\em kinematic trajectories}, $q\in \mathcal{C}^{2n}\left([t_1,t_2], \mathcal{U} \right)\,$, where $\mathcal{U} \subset \RR^m$ is an open subset. Unless clarity demands otherwise, the indices indicating the degree of freedom are suppressed throughout, unless they are necessary for the sake of clarity.

The {\em dynamical trajectory} connecting $(q_1,\ldots q_1^{(n-1)}, t_1)$ and $(q_2,\ldots q_2^{(n-1)}, t_2)$ is the one that makes the action integral stationary while keeping the first \(n-1\) time-derivatives fixed at the boundaries, i.e.\ \(q^{(j)}(t_\alpha)=q^{(j)}_\alpha\) with \(\alpha=1,2\) and \(j=0,\dots,n-1\); namely, it satisfies the variational principle 
\begin{list}
{{\bf (\alph{llista})}}{\usecounter{llista}}
\item $\quad \delta S([q],t_1,t_2) = 0 \,, \qquad $for any variation $\delta q(t)$ such that
\item $\quad  \delta q^{(j)}(t_1)=  \delta q^{(j)}(t_2)= 0 \,, \qquad  j= 0 \ldots n-1\,$.
\end{list}
Similarly as in Section \ref{S2}, the variation reads
\beq \label{en2}
 \delta S \equiv \int_{t_1}^{t_2} \sum_{j=0}^n \frac{\partial L}{\partial q^{(j)}}\, \delta q^{(j)}(t) \,\D t  \,. 
\eeq
but, due to the endpoint conditions {\bf (b)}, the variations of
\(q\) and its first \(n-1\) derivatives vanish at \(t_{1}\) and \(t_{2}\), and
the higher-order variations \(\delta q^{(k)}(t)\) are not independent. Successive integrations by parts of (\ref{en2}) lead to
\beq \label{en3}
 \delta S \equiv \int_{t_1}^{t_2} \D t\, \delta q(t)\,\mathcal{E}\left(q(t), \ldots q^{(2n)}(t), t\right) + 
\sum_{j=0}^{n-1} \left[ p_j(t)\,\delta q^{(j)}(t) \right]_{t_1}^{t_2} \,,
\eeq
where
\begin{eqnarray} \label{en4a}  
 p_{n-1} &:= & \frac{\partial L \left(q, \ldots q^{(n)}, t\right)}{\partial \dot q^{(n)}} \,, \qquad
 p_{j-1} := \frac{\partial L\left(q,\ldots q^{(n)}, t\right)}{\partial q^{(j)}}- \mathbf{D} p_j\,, \quad 1 \leq j < n \,,   \\[2ex]\label{en4b}
\mathcal{E} & := & \frac{\partial L\left(q,\ldots q^{(n)}, t\right)}{\partial q^{(0)}}- \mathbf{D} p_0 \qquad {\rm and} 
 \qquad  \mathbf{D} := \frac{\D \;}{\D t} \equiv \sum_{j} q^{(j+1)}\,\frac{\partial \;}{\partial q^{(j)}} + \frac{\partial \;}{\partial t}  \,,
\end{eqnarray}
which can be iterated to obtain the explicit expressions 
\begin{eqnarray} \label{en4c}
p_{n-1}  &:=& \frac{\partial L }{\partial q^{(n)}} \,, \qquad p_j :=\sum_{l=0}^{n-j-1} (-1)^l \,\mathbf{D}^l \left(\frac{\partial L }{\partial q^{(j+l+1)}} \right)\,, \\[2ex] \label{en4d}
\mathcal{E}  &:=&\sum_{l=0}^{n} (-1)^l \,\mathbf{D}^l \left(\frac{\partial L }{\partial q^{(l)}} \right) \,.
\end{eqnarray}

From the variational identity (\ref{en3}) together with the conditions {\bf (a)} and {\bf (b)}, the stationarity requirement \(\delta S=0\) yields the \(n\)-th-order EL equation
\beq \label{en5}
\mathcal{E}\left(q(t), \ldots q^{(2n)}(t), t\right) = 0\,.
\end{equation}
Assume now that the Lagrangian is \emph{regular}, i.e.\ its Hessian with respect to the highest derivatives is non-degenerate, then the EL system can be reduced to an ordinary differential system.  
By the standard existence and uniqueness theorems for ODEs (see, e.g., Ref.~\cite{Barbu2016, Arnold1992}), there exists a unique solution in the neighborhood of any initial point $(q_1,\ldots q^{(2n-1)}_1, t_1)$; namely, it exists a solution of (\ref{en5}) 
$$\,q(t) = \varphi(t-t_1; q_1,\ldots q^{(2n-1)}_1, t_1)\,, \quad |t-t_1| < T\,, \quad \mbox{such that} \quad q^{(j)}(t_1) = q^{(j)}_1\,,\quad j= 0 \ldots 2n-1\,. $$

As in the first-order setting of Section~\ref{S2}, there is a one-to-one correspondence between initial data $(q^{(0)}_1,\ldots,q^{(2n-1)}_1,t_1)$ and dynamical trajectories. Thus, the set of dynamical trajectories forms a $(2n\,m+1)$-dimensional submanifold within the space of kinematic trajectories $\mathcal{K}$ ---where $m$ denotes the number of degrees of freedom--- and is homeomorphic to the extended initial-data space $\mathcal{D}'=T^{2n-1}\mathcal{U}\times\mathbb{R}$.

Also analogously to the first-order case, if the Lagrangian is a total derivative, it can be written explicitly as
\[
L \equiv \mathbf{D}W \equiv \sum_{l=0}^{n-1} W_l(q,\ldots,q^{(n-1)},t)\,q^{(l+1)} + W_t(q,\ldots,q^{(n-1)},t),\quad\text{with}\quad W_l:=\frac{\partial W}{\partial q^{(l)}}.
\]
A direct calculation then shows that
\[
\frac{\partial L}{\partial q^{(j)}} = \mathbf{D}W_j + W_{j-1},\quad 0\le j\le n,\quad\text{with}\quad W_{-1}=W_n=0.
\]
Substituting these expressions into~(\ref{en4d}) yields $\mathcal{E}\equiv0$. Conversely, if $\mathcal{E}\equiv0$, it follows that the Lagrangian must be a total time derivative. Thus, two Lagrangians produce identical EL equations if and only if their difference is a total derivative.

\subsection{Noether theorem  for higher-order Lagrangian systems \label{S3.1}}
The original formulation of Noether theorem~\cite{Noether} considered, from the very start, field Lagrangians depending on derivatives of a finite arbitrary order. For a mechanical
system, the derivation proceeds exactly as in Section~\ref{S2.1}, \textit{mutatis mutandis}, with the obvious modifications to accommodate the higher derivatives.
Under an infinitesimal transformation of the form~(\ref{e6}), the \(n\)-th derivative of the coordinate transforms as
\begin{equation}  \label{en6a}
q^{\prime (n)} (t^\prime)=  \left( \left[1 - \delta \dot T(t)\right]\,\frac{\D\;}{\D t}\right)^n\,q(t) + \delta q^{(n)} (t) \,,
\end{equation}
while a higher-order Lagrangian \(L(q^{(j)},t)=L(q,\dot q,\ldots,q^{(n)},t)\) changes according to
\begin{equation}  \label{en7}
L^\prime\left(q^\prime(t^\prime),\ldots q^{\prime\,(n)} (t^\prime), t^\prime\right) = \frac{\D t}{\D t^\prime}\,L\left(q(t), \ldots q^{(n)}(t),t \right) \,,
\end{equation} 
so that the action integral remains invariant for any kinematic trajectory.

Under a {\em Noether symmetry}, the Lagrangian may change only by a total
time derivative,
\begin{equation}  \label{en7b}
L^\prime\left(q^\prime,\ldots q^{\prime\,(n)}, t^\prime\right) = L \left(q^\prime,\ldots q^{\prime\,(n)}, t^\prime\right)+ \frac{\D W\left(q^\prime,\ldots q^{\prime\,(n-1)}, t^\prime\right)}{\D t^\prime}  \,,
\end{equation} 
and therefore the EL equations (\ref{en4d}) remain invariant under the transformation. In such a case, the condition that the action integral is invariant for any kinematic trajectory is the analogous of (\ref{e7a}) 
$$  \int_{t^\prime_1}^{t^\prime_2} \D t^\prime\,L^\prime\left(q^\prime(t^\prime),\ldots q^{\prime\,(n)} (t^\prime), t^\prime\right)  \equiv \int_{t_1}^{t_2} \D t\, L\left(q(t), \ldots q^{(n)}(t),t \right) \, 
$$
that, combined with~\eqref{en7b}, yields    
\begin{equation*}  
\int_{t_1}^{t_2} \D t\,\left\{ L\left(q^{\prime\,(j)}(t),t) \right) -  L\left( q^{(l)}(t),t \right)+ 
\frac{\D }{\D t} \left[ W\left(q(t),\ldots q^{(n-1)}, t\right) + L \,\delta T \right]  \right\}   \equiv 0 \,
\end{equation*}
for arbitrary $[t_1,t_2]\,$.  Hence
$$ L\left(q^{\prime\,(j)}(t),t) \right) -  L\left(q^{(l)}(t),t \right)+ \frac{\D }{\D t} \left[ 
W\left(q(t),\ldots q^{(n-1)}, t\right) + L \,\delta T \right]     \equiv 0 \,. $$
Proceeding as in Section \ref{S2.1} and using (\ref{e6}) and (\ref{e6a}), we can write the latter equation as
$$ \sum_{j=0}^n \frac{\partial L}{\partial q^{(j)}}\,\tilde\delta  q^{(j)} + \frac{\D (L\,\delta T + W)}{\D t} \equiv 0 \,,$$
where $\,\tilde\delta  q^{(j)}\,$ denotes the $j$-th time derivative of 
$\,\tilde\delta  q := q^\prime(t) - q(t) = \delta q(t)- \dot q(t)\,\delta T(t)\,$.
 
Then, including  (\ref{en4a}) and (\ref{en4b}), the above equation becomes
$$ \left(\mathcal{E} + \frac{\D p_0}{\D t} \right)\,\tilde\delta q^{(0)} + \sum_{j=1}^{n-1} \left( p_{j-1} + \frac{\D p_j}{\D t} \right)\,\frac{\D \tilde\delta q^{(j-1)}}{\D t} + p_{n-1}\,\,\frac{\D \tilde\delta q^{(n-1)}}{\D t} + \frac{\D (L\,\delta T + W)}{\D t} \equiv 0 \, $$
and, after rearranging and applying Leibniz' rule, we arrive at the \emph{Noether identity}
\beq  \label{en8a}
\mathcal{E}(q)\,\tilde\delta q^{(0)} + \frac{\D }{\D t} \left[ \sum_{j=0}^{n-1}  p_{j}\,\tilde\delta q^{(j)} + L\,\delta T + W \right] \equiv 0 \,.
\eeq
Consequently, the quantity
\beq  \label{en8b}
Q :=  \sum_{j=0}^{n-1}  p_{j}\,\tilde\delta q^{(j)} + L\,\delta T + W 
\eeq
is conserved along any dynamical trajectory (\(\mathcal{E}=0\)).

Using the binomial identity, we have that
$$  \tilde\delta q^{(j)} \equiv \delta q^{(j)} - \frac{\D^j (\dot q \,\delta T)}{\D t^j} \equiv \delta q^{(j)} - \sum_{l=0}^j  {j \choose l} 
\,\delta T^{(l)}\,q^{(j-l+1)} \,. $$  %
We can then rewrite the conserved quantity~\eqref{en8b} in the more explicit
form
\beq  \label{en8c}
Q:= W + \sum_{j=0}^{n-1}  p_{j}\,\delta q^{(j)} - E\,\delta T + \sum_{l=1}^{n-1} \delta T^{(l)} \, \sum_{j=l}^{n-1}  {j \choose l}  \, p_j\,q^{(j-l+1)} \,,
\eeq
where 
\beq  \label{en8d}
E:= \sum_{j=0}^{n-1}  p_{j}\, q^{(j+1)} - L
\eeq
is the Ostrogradski energy \cite{Ostrogradski,Whittaker}. When \(\delta T\) is a constant, the last term in~\eqref{en8c} drops out. In particular, for a pure time translation \(\delta T=-\varepsilon\) with \(\delta q=0\), the conserved Noether charge is the energy~\(E\).

\subsection{Hamiltonian formalism \label{S3.2}}
The construction follows the same steps as in Section~\ref{S2.2} for first-order systems. The prefactor of $\delta q^{(j)}\,, \;\, j= 0 \ldots n-1\,$, in (\ref{en8c}) defines the canonical (Ostrogradski) momentum 
\beq  \label{en9}
p_j(q,\ldots q^{(2n-1)}, t) = \sum_{l=0}^{n-j-1} (-1)^l \,\frac{\D^l \;}{\D t^l} \left(\frac{\partial L }{\partial q^{(j+l+1)}} \right) \,,
\eeq
on which the Legendre-Ostrogradski  transformation is based, and $E(q,\ldots q^{(2n-1)}, t)$ is the energy.

Following the pattern of Section~\ref{S2.2}, we introduce the time-dependent $2$-form
\beq  \label{en9a}
   \beta :=\omega  - \D E \wedge \D t\,, \qquad {\rm with} \qquad \omega:= \sum_{l=0}^{n-1} \D p_l (q,\ldots q^{(2n-1)},t)\wedge \D q^{(l)}  \, 
\eeq
and the generator of time evolution, namely, the total time derivative
$$  \mathbf{D} := \frac{\D \;}{\D t} = \sum_{j=0}^{2n-1} q^{(j+1)}\,\frac{\partial \;}{\partial q^{(j)}} + \partial_t   \,.$$ 
The interior product of $\mathbf{D}$ with $\beta$ is
\begin{eqnarray*}
i_\mathbf{D} \beta & \equiv & \sum_{l=0}^{n-1}\left(\mathbf{D} p_l\,\D q^{(l)} - q^{(l+1)}\,\D p_l \right) - \mathbf{D} E \,\D t + \D E  
\nonumber \\[1.5ex]
   & \equiv & \sum_{l=0}^{n-1}\left(\mathbf{D} p_l\,\D q^{(l)} - q^{(l+1)}\,\D p_l \right)- \mathbf{D}E \,\D t + 
	\sum_{l=0}^{n-1}\left( p_l \,\D q^{(l+1)} -  q^{(l+1)} \,\D p_l \right) - \D L           \nonumber\\[1.5ex] 
  & \equiv & - \left(\mathbf{D} E + \partial_t L \right)\,\D t - \mathcal{E}\,\D q^{(0)} 	\,,
\end{eqnarray*}
where Eqs.~\eqref{en4b}-\eqref{en4c} have been used in the last step. Furthermore, one shows directly that
$$ \mathbf{D} E + \partial_t L \equiv - \mathcal{E}\, q^{(1)} \,$$
so that the previous result yields
\beq  \label{en9b}
   i_\mathbf{D} \beta \equiv - \mathcal{E} \,\left(\D q^{(0)} -  q^{(1)}\,\D t \right) \,,
\eeq
where $\,\mathcal{E}\,$ is the EL equation. Consequently,
\beq  \label{en9c}
   i_\mathbf{D} \beta = 0 \qquad \quad \Leftrightarrow \qquad \quad \mathcal{E} = 0 \,, 
\eeq
i.e.\ the EL equation is equivalent to the intrinsic equation
\beq  \label{en9d}
  i_\mathbf{D} \beta = 0 \,.
\eeq

So far ---exactly as in Section~\ref{S2.2} for first-order dynamics--- we have
not assumed that the Lagrangian is \emph{regular}. When regularity holds, the Legendre-Ostrogradski map~\eqref{en9} is locally invertible and can be solved for the derivatives higher than $n$,  
$$ q^{(\alpha)} = q^{(\alpha)}\left(q^{(0)} \ldots q^{(n-1)}, p_0 \ldots p_{n-1}, t \right)  \,,\qquad \alpha = n ,\ldots 2n-1 $$
Hence $\,\left(q^{(j)}, p_l, t \right)_{j,l = 0 \ldots n-1}  \,$ furnish independent canonical coordinates on the extended phase space.
With these coordinates, the canonical 2-form $\,\omega:= \D p \wedge \D q\,$ is symplectic on the cotangent bundle $\,T^\ast\left(T^{n-1}\mathcal{U}\right)\,,$ while $\beta$ provides a contact form \cite{Godbillon} on the extended phase space $\,T^\ast\left(T^{n-1}\mathcal{U}\right) \times \RR\,$.  The condition \(i_{\mathbf D}\beta=0\) then determines a \emph{unique} solution $\mathbf{D}\,$.

Defining the Hamiltonian as $\,H\left(q^{(j)}, p_l, t \right) := E \left(q^{(j)}, q^{(\alpha)}(q^{(j)}, p_l, t),t\right)\,,$ where \(E\) is the Ostrogradski energy~\eqref{en8d}, and substituting in~\eqref{en9d}, we obtain
$$ i_\mathbf{D}\omega + \D H + \partial_t L\,\D t = 0 \,, $$ 
whose components yield the Hamilton-Ostrogradski equations
for \(j=0,\ldots,n-1\): 
\beq  \label{en9e}
\frac{\D q^{(j)}}{\D t} = \frac{\partial H}{\partial p_j} \,, \qquad \frac{\D p_l}{\D t} = - \frac{\partial H}{\partial q^{(l)}}  \,, \qquad \mbox{and the well-known relation} \qquad 
 \frac{\partial H}{\partial t} = - \frac{\partial L}{\partial t} \,.
\eeq

\section{Infinite-order Lagrangians  \label{S4} }
We have already outlined in the Introduction how infinite-order Lagrangians began to appear in physics. Indeed, attempts to control ultraviolet divergences typically lead to Lagrangians containing nonlocal-path-dependent-terms rather than infinite-order ones. However, since the application of variational principles and standard Lagrangian mechanics methods to nonlocal Lagrangians was not initially clear, Marnelius \cite{Marnelius1973} developed a procedure to transform these nonlocal Lagrangians into infinite-order ones. He achieved this by "\ldots making use of the variation method for Lagrangians with higher-order time derivatives \underline{formally} generalized to infinite-order Lagrangians" (emphasis ours).

The term ``formally" here refers explicitly to the simple replacement of a finite order $n$ with infinity ($\infty$)\footnote{Finite sums are replaced with series that must be handled formally because their convergence is uncertain.}. This approach is heuristic, relying on the use of Taylor series expansions, which are not necessarily convergent. Typically, this yields a set of series that Marnelius managed to sum formally \cite{Marnelius1973}, producing nonlocal expressions usually expressed through integral operators. Given the heuristic nature of this approach, any results derived must eventually be verified explicitly to ensure their validity ---such as confirming whether an integral of motion proposed through this method truly holds.

To be more specific, a nonlocal Lagrangian is a functional $\mathcal{L}(q)$, depending on the entire trajectory $q(\tau)$ (a smooth function), or at least a bounded segment of it, rather than solely on the value of the coordinate and a finite number of its derivatives at a single instant $t$. For simplicity, we consider only time-independent Lagrangians with a single degree of freedom. Extending to multiple degrees of freedom presents no conceptual difficulties. For clarity and simplicity of notation, we avoid writing $[q]$, typically used to emphasize functional dependence. Usually, nonlocal Lagrangians involve operations over trajectory parameters, for example:
$$  \mathcal{L}\left(q\right) = \int_{\RR^2} \D\tau\,\D\sigma\,K(\tau,\sigma)\,q(\tau)\,q(\sigma) \,. $$
In turn, an infinite-order Lagrangian is a function $L(q^{(j)}) = L(q, \dot q, \ldots, q^{(n)}, \ldots)$ defined on an open subset $\mathcal{U} \subset \mathbb{R}^\mathbb{N}$, where $q^{(j)}$ denotes the $j$-th order derivative.

A formal correspondence connecting $\mathcal{U}$ and the space of smooth trajectories $\mathcal{C}^\infty (\mathbb{R})$ can be established by employing the formal Taylor expansion\footnote{$T_t q$ represents ``the same'' trajectory as $q$, but with its ``initial point'' shifted from $q(0)$ to $q(t)$.}:  
\beq  \label{ei2}
T_t q(\tau) := q(t+\tau) \equiv \sum_{j=0}^\infty \frac{\tau^j}{j!}\, q^{(j)}(t) \,.
\eeq
This provides a translation rule $T_t q \leftrightarrow \left(q^{(j)}(t)\right)$, which, when substituted into $\mathcal{L}(T_t q)$ and evaluated through the operations involved in the definition of $\mathcal{L}$ ---usually integrals--- yields:
\beq  \label{ei3a}
\mathcal{L}\left(T_t q\right)  \longleftrightarrow L\left( q^{(j)}(t)\right) \,.
\eeq
This correspondence underpins the entire construction developed by Marnelius \cite{Marnelius1973}. However, it is important to acknowledge the limitations and inaccuracies associated with this reasoning, namely:

\begin{list}{(\alph{llista})}{\usecounter{llista}}
\item Expression (\ref{ei2}) is not valid for general smooth trajectories but only for real-analytic ones with infinite convergence radius.
  
\item All derivatives $q^{(j)}(t)$ at a particular instant $t$ do not uniquely determine a smooth function $T_t q(\tau)$. According to Borel's theorem \cite{Borel}, there exists at least one smooth function matching these derivatives, but this function is not unique ---in fact, infinitely many smooth functions vanish at a given point along with all their derivatives.

\item The correspondence defined by (\ref{ei2}) introduces an additional issue obstructing the direct application of the variational principle. Indeed, if  $\delta q(t_1)$ and all its derivatives vanish at a point $t_1$, then (\ref{ei2}) implies that $\delta q(t_1+\tau)=0$,  and consequently, $\delta q(t)$ and all its derivatives vanish at every point $t$. This is closely related to a subtle issue regarding endpoint conditions, which we analyze thoroughly in Section \ref{Paradox}.
\end{list}

Despite these limitations, we present below the results obtained from this approach. The action integral associated with an infinite-order Lagrangian is given by
\beq  \label{ei1}
S([q],t_1,t_2) := \int_{t_1}^{t_2} L\left( q^{(j)}(t)\right)\,\D t  \,.
\eeq  
As stated earlier, Marnelius' proposal involves exploiting the variational principle as if the Lagrangian were of finite order $n$ and subsequently replacing $n$ by infinity ($\infty$). Accordingly, the time evolution generator (\ref{en4b}), the variation of the action (\ref{en3}), the canonical momenta (\ref{en4c}), and the EL equation (\ref{en4d}) become, respectively:
\begin{eqnarray} \label{ei4z}
 \mathbf{D} &:=& \partial_t + \sum_{j=0}^\infty q^{(j+1)}\,\frac{\partial \;}{\partial q^{(j)}}  \,, \\[2ex] \label{ei3}
\delta S &\equiv& \int_{t_1}^{t_2} \D t\, \delta q(t)\,\mathcal{E}\left(q^{(j)}(t)\right) + \sum_{j=0}^\infty \left[ p_j\left(q^{(k)}(t)\right)\,\delta q^{(j)}(t) \right]_{t_1}^{t_2} \,, \\[2ex]  \label{ei4c}
 p_j\left(q^{(k)}(t)\right)  & := & \sum_{l=0}^\infty (-1)^l\,\mathbf{D}^l\left(\frac{\partial L}{\partial q^{(j+l+1)}}\right)_{\left(q^{(k)}(t)\right)} \,,\qquad j= 0, 1, 2, \ldots \quad {\rm and}\\[2ex] \label{ei4}
\mathcal{E}\left(q^{(k)}(t)\right)  &:= & \sum_{l=0}^\infty (-1)^l \,\mathbf{D}^l \left(\frac{\partial L }{\partial q^{(l)}} \right)_{\left(q^{(k)}(t)\right)}= 0\;.
\end{eqnarray}
Recall that, in the standard notation of classical mechanics, it is typically understood that
\[
\left( \frac{\partial L}{\partial q^{(l)}} \right)_{\left(q^{(k)}(t)\right)} 
:= \frac{\partial L\left(q(t), \ldots, q^{(n)}(t), \ldots\right)}{\partial q^{(l)}}\,.
\]
Note that equations (\ref{ei4c}) and (\ref{ei4}) imply, respectively, that
\beq \label{ei4y}
 \mathbf{D} p_l + p_{l-1} - \frac{\partial L}{\partial q^{(l)}} \equiv 0\,, \qquad l \geq 1 \,,\qquad {\rm and}\qquad
\frac{\partial L}{\partial q^{(0)}}  -\mathbf{D} p_0 \equiv \mathcal{E} \,,
\eeq
which will be useful later.

As in the finite-order cases discussed earlier, a Lagrangian that is a total time derivative, $L\left(q^{(k)}\right) = \mathbf{D} W\left(q^{(k)}\right),$ yields an identically vanishing EL equation. Indeed, equation~(\ref{ei4}) becomes:
\[
\mathcal{E} \equiv \sum_{l=0}^{\infty} (-1)^l\, \mathbf{D}^l \left( \frac{\partial}{\partial q^{(l)}}\, \mathbf{D}W \right),
\]
which, upon using the commutator identity
\[
\left[ \frac{\partial}{\partial q^{(l)}}, \mathbf{D} \right] = \sum_{j=0}^\infty \left[ \frac{\partial}{\partial q^{(l)}}, q^{(j+1)} \frac{\partial}{\partial q^{(j)}} \right] = \frac{\partial}{\partial q^{(l-1)}}, \qquad \text{for } l \geq 1,
\]
and $\displaystyle{\left[ \frac{\partial}{\partial q^{(0)}}, \mathbf{D} \right] = 0},$
becomes:
\[
\mathcal{E} \equiv \sum_{l=0}^{\infty} (-1)^l\, \mathbf{D}^{l+1} \left( \frac{\partial W}{\partial q^{(l)}} \right)
+ \sum_{l=1}^{\infty} (-1)^l\, \mathbf{D}^{l} \left( \frac{\partial W}{\partial q^{(l-1)}} \right).
\]
Replacing the index in the second formal series via \( l = k - 1 \), with \( k \geq 1 \), we find that the two series cancel term by term, yielding
\[
\mathcal{E} \equiv 0.
\]
However, proving the converse ---namely, that if the EL equation vanishes identically, then the Lagrangian must be a total time derivative--- is not so straightforward as in the finite-order case. In fact, such a statement may not hold in general for infinite-order Lagrangians.

\subsection{The phase space \label{PhSp}} 
So far the parallel with finite-order Lagrangian systems. However, some important features are lost along the way. To begin with, examining the equation of motion: 
\beq \label{ei4aa}
\mathcal{E}\left(q^{(0)}, q^{(1)}, \ldots q^{(n)},\ldots \right)  = 0\,,
\eeq
we immediately notice the absence of a highest-order derivative. In contrast to equation (\ref{en4d}), it cannot be transformed into an ordinary differential system (ODS) in normal form. Consequently, standard existence and uniqueness theorems no longer apply, and we must adopt an alternative perspective.

For finite-order ($n$) systems, standard existence and uniqueness theorems establish a one-to-one correspondence between each solution $q(t+\tau)$ and a finite set of initial conditions, namely $\left(q(t), \dot q(t), \ldots q^{(2n-1)}(t) \right)\,$. If we continue following the rule of replacing $n$ infinity  ($\infty$), the initial conditions for an infinite-order system become a sequence $\left(q(t), \dot q(t), \ldots q^{(j)}(t),\ldots \right)\,$. However, once this infinite initial sequence is provided, the time evolution is already determined entirely by the Taylor series (\ref{ei2}). Thus, equation (\ref{ei4aa}) should not be interpreted as an ordinary differential equation governing the temporal evolution $q(t+\tau)\,$,  but rather as a constraint on the initial-data sequence. Furthermore, since (\ref{ei4aa}) must hold true for all times $t$, all of its time derivatives must also be satisfied: 
\beq \label{ei4a}
\mathcal{E}^{(l)}\left(q^{(0)}, q^{(1)}, \ldots q^{(n)},\ldots \right) :=  \mathbf{D}^l\,\mathcal{E}\left(q^{(0)}, q^{(1)}, \ldots q^{(n)},\ldots \right) = 0 \,,\qquad l= 0, 1, 2 \ldots \,.
\eeq
These conditions generate an infinite set of secondary constraints, which can be solved to express the infinite set of initial derivatives $q^{(l)}(t)\,$ in terms of a subset of them, finite or infinite in number. The \textit{phase space} is then identified as the set of all initial sequences that fulfill these constraints (\ref{ei4a}).

It is straightforward to verify that this approach can also be applied to regular $n$-th order Lagrangian systems, reproducing the familiar results presented in Section \ref{S3}. Indeed, the EL equation can be expressed in normal form as: 
$$ q^{(2n)} = f_{2n} \left(q,\dot q, \ldots  q^{(2n-1)}\right)  \,, $$
and their time derivative is then given by:
$$  q^{(2n+1)} = \sum_{j=0}^{2n-2} q^{(j+1)}\,\frac{\partial f_{2n}}{\partial q^{(j)}} + \frac{\partial f_{2n}}{\partial q^{(2n-1)}} \, f_{2n}\left(q,\dot q, \ldots  q^{(2n-1)}\right) =: f_{2n+1}\left(q,\dot q, \ldots  q^{(2n-1)}\right)  \,. $$
Iterating this procedure, all derivatives of order higher than $2n$ can be expressed in terms of lower-order derivatives. Thus, in the regular finite-order ($n$) case, the space of initial sequences ---namely, the phase space--- is a  $2n$-dimensional manifold, coordinatized naturally by $q$ and its derivatives up to order $2n-1\,$.

In the infinite-order case, no general rule exists for explicitly obtaining a coordinate system for the submanifold of initial sequences satisfying the constraints (\ref{ei4a}). Consequently, the best available approach is to handle the phase space in implicit form. Identifying a suitable coordinate system must be addressed individually for each specific case, and it is not guaranteed that a finite number of parameters will always be sufficient.

\subsubsection{Some examples \label{PhSpEx}}
The aim of the following examples is to show how one can identify a minimal set of parameters that uniquely characterize each solution of equation (\ref{ei4}), thereby providing a coordinate system for the phase space. Consider the Lagrangian
\beq \label{ei30}
L = \frac12\,q\,F(D) q\,, 
\eeq
where $D$ is the total time derivative defined in (\ref{ei4z}) and 
\beq \label{ei30a}
F(z) = \sum_{j=0}^\infty f_j\,z^j \,. 
\eeq
is a holomorphic function. One easily checks that the EL equation (\ref{ei4}) becomes
\beq \label{ei30b}
F_+(D) q = 0\,, \qquad {\rm where} \quad F_+(z) = \frac12\,\left[ F(z) + F(-z) \right]  \,.
\eeq
Two cases then arise, depending on whether the number $n$ of zeros of $F_+(z)\,$ is finite or infinite\footnote{See Ref.\cite{Barnaby} for a similar approach.}. \\
\begin{description}
\item[Finite case \cite{Vladimirov}:] If $n$ is finite, $F_+(z)$ factors as: $F_+(z) = P(z)\,\gamma(z)\,$, where $P(z)$ is a degree-$n$ polynomial whose roots $\{\alpha_j\}$ have multiplicities $\{r_j\}$, and $\gamma(z)$ is a nonvanishing holomorphic function. Equation (\ref{ei30b}) then reads $\,P(D) \,\gamma(D) q = 0\,,$ which implies
\beq \label{ei31}
\gamma(D) q = \psi\,, \qquad {\rm with} \qquad \psi(t) = \sum_j Q_j(t) \, e^{\alpha_j t} \,,
\eeq
where each $ Q_j(t) $ is a polynomial of degree $r_j - 1\,$, in accord with the theory of constant-coefficient linear ODEs.  

We seek solutions of the form
$$ q(t) = \sum_{j=1}^k q_j(t)\,e^{\alpha_j t} \,, \qquad \text{with} \qquad  \gamma(D)\left(q_j(t)\,e^{\alpha_j t}\right)  = Q_j(t)\,e^{\alpha_j t} \,. $$
Since $\, \gamma(D)\left(u(t)\,e^{\alpha t}\right) = e^{\alpha t}\,\gamma(D+\alpha) u(t)\,$, we obtain
$\, \gamma\left(D+\alpha_j\right) q_j(t) = Q_j(t) \,.$ Taking Fourier transforms (``hats'' denote the FT) gives 
$$ \gamma\left(-i\omega+\alpha_j\right) \hat{q}_j(\omega) = \hat{Q}_j(\omega)  \,,$$
and because $\gamma$ has no zeros, the latter equation can be solved in $\hat{q}_j(\omega)$. Inverting the FT and using the convolution theorem yields
$$  q_j(t) =\int_\RR \D\tau \,G_j(t-\tau)\,Q_j(\tau) \,, \qquad {\rm with} \qquad G_j(t) =\frac1{2\,\pi}\,\int_\RR \D\omega\,
\frac{e^{-i \omega t}}{\gamma(-i\omega+\alpha_j)}  \,. $$
Writing the polynomial as $\displaystyle{Q_j(\tau) = \sum_{l=0}^{r_j-1} a_{jl} \tau^l }\,$ gives
$$q_j(t) = \sum_{l=0}^{r_j-1} a_{jl} \,G_j^l(t) \,, \qquad {\rm with} \qquad G_j^l(t)= \int_\RR \D\tau\,G_j(t-\tau)\,\tau^l \,. $$
Hence the solution $q(t)$ depends on the $n=r_1 + \ldots + r_k\,$ coefficients of the polynomials $Q_j(\tau)\,$, which are connected with the values of $q(t)$ and its derivatives up to the order $n-1$ at $t=0\,$.  In other words, the phase space is finite-dimensional.

\item[Infinite case:] An example of $F_+(z)$ with an infinite number of zeros is the genuinely nonlocal Pais-Uhlenbeck oscillator \cite{PaisUhlenbeck1950},  defined by 
$$ F(z) = \prod_{n=1}^\infty \left(1 + \frac{z^2}{\omega_n^2} \right) \,, \qquad \omega_ 1 < \omega_2 < \ldots < \omega_n < \ldots \,.$$
In the special case  $\,\omega_n = \pi\,n\,,$ the infinite product can be calculated and it yields
\beq \label{ei40}
 F(z) = \frac{\sinh z}{z} = \sum_{n=0}^\infty \frac{z^{2n}}{(2 n + 1)!} \,.
\end{equation}
The EL equation (\ref{ei30b}) then becomes
\beq \label{ei40a}
  F(D) q(t) \equiv \prod_{n=1}^\infty \left(1 + \frac{D^2}{\omega_n^2} \right)\, q = 0 \, , 
\end{equation} 
whose general solution is a superposition of the individual harmonic modes satisfying  $\;\left(\pi^2 n^2 + D^2\right)\,q = 0 \,$, namely:
\beq \label{ei40b}
q(t) = \sum_{n=1}^\infty \left(A_n \,e^{i \pi n t} + A^\ast_n \,e^{-i \pi n t} \right) \,, \qquad A_0=0 \,,
\end{equation}
where the fact that $q(t)$ is real has been included. 

To verify that this is indeed the general solution, we include (\ref{ei40}) to write the rhs of (\ref{ei40a}) and perform the formal addition
$$ F(D) q(t) \equiv \sum_{l=0}^\infty \frac{q^{(2l)}(t)}{(2 l + 1)!} \equiv \sum_{n=0}^\infty \frac{q^{(n-1)}(t)}{n!} \,\frac{\left[1- (-1)^n\right]}{2} \,$$
that can be put as
$$ F(D) q(t) \equiv \frac12\,\left[ \sum_{n=0}^\infty \frac{Q^{(n)}(t)}{n!} -  \sum_{n=0}^\infty \frac{Q^{(n)}(t)}{n!}\,(-1)^n \right] \equiv 
\frac12\,\left[ Q(t+1) - Q(t-1) \right] \,,$$
where $Q(t)$ is a primitive of $q(t)\,.$ Therefore, the EL equation becomes
\beq \label{ei41}
 F(D) q(t) \equiv \frac12\,\int_{t-1}^{t+1} \D\tau\,q(\tau) = 0 \,, \qquad \forall t\in \RR \,.
\end{equation}
Particularizing this equation at $t=0$ and considering its first derivative, we conclude that: \\[1ex]
\centerline{$q$ is a periodic function with period 2 and vanishing mean value.}

\noindent
Being $q$ smooth, the Fourier series theorem \cite{Apostol} implies that the expansion (\ref{ei40b}) yields the general solution. 
Hence, the phase space can be coordinatized either by the set of complex sequences $\,\left\{ A_n,  \, n \in \NN \right\}\,$ or by the set of functions $\, q \in \mathcal{C}^0 ([-1,1])\,$ with vanishing mean value and fulfilling the constraint $q(-1)=q(1)\,$.
\end{description}

\subsection{Noether theorem  for  infinite-order Lagrangians \label{S4.1}}
Replacing $n$ with $\infty$ in the Noether identity (\ref{en8a}) yields
$$ \mathcal{E}\,\tilde\delta q^{(0)} + \mathbf{D} \left[ \sum_{j=0}^\infty p_{j}\,\tilde\delta q^{(j)} + L\,\delta T + W \right] \equiv 0 \,,$$
where $ \tilde\delta q (t) = \delta q (t) - \dot q(t) \,\delta T(t)\,$.
Focusing on time translations $\delta T = {\rm constant}\,,$ this simplifies to
\beq  \label{ei8a}
\mathcal{E}\,\tilde\delta q^{(0)} + \mathbf{D} \left[ \sum_{l=0}^\infty p_{l}\,\delta q^{(l)} -E\,\delta T + W \right] \equiv 0 \,,
\eeq
where the prefactor $p_l$ of $\delta q^{(l)}$ is given by (\ref{ei4c}) and 
\beq  \label{ei8b}
E = \sum_{l=0}^\infty p_{l}\, q^{(l+1)} - L \,
\eeq
is the energy. Hence the quantity
\beq  \label{ei8c}
Q :=\sum_{l=0}^\infty p_{l}\,\delta q^{(l)} -E\,\delta T + W  
\eeq
is conserved along all dynamical trajectories. It is worth noting that the first term on the right-hand side coincides with the boundary contribution on the right-hand side of equation~(\ref{ei3}), just as in the case of finite-order Lagrangians discussed in Sections~\ref{S2} and~\ref{S3} ---an outcome that is to be expected.

\subsection{Hamiltonian formalism  \label{S4.2}}
We begin by mimicking the finite-order construction, replacing all finite sums with their formal infinite-series counterparts. As in Sections \ref{S2.2} and \ref{S3.2}, we introduce the differential two-form
$$\,\beta :=\omega  - \D E \wedge \D t\,,$$ 
where $E$ is the energy defined in (\ref{ei10a}), $\,\mathbf{D}\,$ is the generator of time evolution (\ref{ei4z}) and
\beq \label{ei11a}
   \omega:= \sum_{l=0}^\infty \D p_l(q^{(j)}) \wedge \D q^{(l)}  \,  
\eeq
is the presymplectic form with the function $\,p_l(q^{(j)})\,$ being the momenta (\ref{ei4c}) ---or, what is the same, the prefactors of $\delta q^{(l)}$ in the Noether identity (\ref{ei8a}). One then checks straightforwardly that $\beta$ is closed.

Following the same steps as in Sections \ref{S2.2} and \ref{S3.2} but replacing $n$ with $\infty\,$, one finds
\beq  \label{ei9b}
   i_\mathbf{D} \beta \equiv - \mathcal{E} \,\left(\D q^{(0)} - q^{(1)}\,\D t \right) \,,
\eeq
so that
\beq  \label{ei9c}
   i_\mathbf{D} \beta = 0 \qquad \quad \Leftrightarrow \qquad \quad \mathcal{E} = 0 \,.
\eeq
In other words, the infinite-order EL equation (\ref{ei4}) is exactly equivalent to the presymplectic condition:
\beq  \label{ei9d}
  i_\mathbf{D} \beta = 0 \,.
\eeq

As in the finite-order treatments (Sections \ref{S2.2} and \ref{S3.2}), no regularity assumption on the Lagrangian is required to derive equation (\ref{ei9d}). In fact, regularity ---which refers to the non-degeneracy of the Hessian of highest-order velocity--- becomes ill-posed for infinite-order Lagrangians, since there is no ``highest'' velocity. Consequently, the final steps in Sections \ref{S2.2} and \ref{S3.2} that yield the standard Hamilton equations have no analogue in the infinite-order setting.

The presymplectic form $\omega $ is a closed 2-form on the kinematic space $\mathcal{K}\,$, whose coordinates are $\,\left(q^{(0)},q^{(1)}, \ldots q^{(l)} \ldots \right)\,$. 
The next step toward a Hamiltonian formulation is to identify a maximally independent subset of these coordinates $\left(q^{(j)}\right)_{j=0, 1, \ldots }\,$. While there is no universal prescription, inspecting the structure of  (\ref{ei11a}) often reveals which combinations to choose, as the following examples will illustrate.

\begin{description}
\item[A simple degenerate case:] Consider a regular first-order Lagrangian $\, L\left(q^{(0)},q^{(1)}\right)\,$. From (\ref{ei4c}), the only nonzero momentum is
\begin{equation}  \label{ei10}
 p_0 = \frac{\partial L\left(q^{(0)},q^{(1)}\right)}{\partial q^{(1)}} \,, \qquad p_j \equiv 0 \,, \quad j \geq 1 \,,
\end{equation}
and the energy (\ref{ei10a}) becomes $\, E = p_0 \,q^{(1)} - L(q^{(0)},q^{(1)})\,$. Since $p_j \equiv 0$ for all $j\geq 1$, viewing $L$ as an infinite-order Lagrangian makes it manifestly degenerate.

The presymplectic form (\ref{ei11a}) reduces to
$$ \omega = \D p_0(q^{(0)},q^{(1)}) \wedge \D q^{(0)} \in \Lambda^2\left(\mathcal{K} \right) $$
depending only on $q^{(0)}\,$ and $\,q^{(1)}\,$. Hence this suggests that the phase space for a regular first-order Lagrangian system is a submanifold of the space of infinite sequences $\RR^\NN $, which is coordinated by $\,q^{(0)}\,$ and $\,p_0\left(q^{(0)},q^{(1)} \right)\,$. The implicit equations of this submanifold are the EL equation (\ref{ei4aa}) together with all its time derivatives (\ref{ei4a}), which allow to obtain  $q^{(j)}\,, \;\, j \geq 2\,$ in terms of  $q^{(0)}\,$ and $q^{(1)}\,$.
Something similar holds for higher-(finite)-order Lagrangians.
\end{description}

\begin{description}
\item[An infinite order case:] Consider the Lagrangian of the genuinely nonlocal Pais-Uhlenbeck oscillator \cite{PaisUhlenbeck1950}:
\begin{equation}  \label{ei11z}
 L = \frac12\,q^{(0)}\, \sum_{n=0}^\infty \frac{q^{(2n)}}{(2n+1)!}  \,.
\end{equation}
From the momentum formula (\ref{ei4c}), one finds
\begin{equation}  \label{ei12}
 p_j =  \sum_{n,l=0}^\infty \frac{(-1)^l q^{(l)}}{2 (l+j+2)!}\,\delta^{j+l+1}_{2n} \,.
\end{equation}
Substituting into the presymplectic form (\ref{ei11a}) gives
$$ \omega = -\sum_{k,m=0}^\infty \frac1{(2k+2m+3)!}\,\D q^{(2k+1)} \wedge \D q^{(2m)} \,. $$ One can then introduce canonical coordinates by setting
\begin{equation}   \label{ei13a}
Q^m = q^{(2m)} \,,\qquad P_m= \sum_{k=0}^\infty A_{m k}\,q^{(2k+1)}\,, \quad
A_{m k} = - \frac1{(2k+2m+3)!} \,,
\end{equation}
so that
\begin{equation}   \label{ei13}
\omega = \sum_{m=0}^\infty \D P_m \wedge \D Q^m \,.
\end{equation}
Here the the canonical coordinates $Q^m$ are the even derivatives of $q$, while the canonical momenta $P_m$ depend on the odd derivatives. In principle, one then inverts the infinite matrix $A_{mk}$ (i.e. finds $A^{km}$ with $\sum_{m} A^{km}A_{ml} = \delta^k_l$) to express all odd derivatives in terms of the momenta: 
$$q^{(2k+1)} = \sum_{m=0}^\infty A^{k m}\, P_m\,.$$ Substituting into the energy (\ref{ei8b}) and simplifying yields the Hamiltonian
\begin{equation}  \label{ei14}
 H = \frac12 \,\sum_{m,k=0}^\infty A^{k m}\,P_k\,P_m + \frac12\,\left[\sum_{m,k=1}^\infty \frac{Q^k\,Q^m}{(2k+2m+1)!} -Q^0\,Q^0\right]\,.
\end{equation} 
Although manipulating these infinite sums directly is cumbersome, the fully nonlocal formalism presented in Section \ref{S5} makes this process considerably more tractable.
\end{description}

\subsection*{Lagrange equations in functional form  \label{S4.0} }

Marnelius' procedure \cite{Marnelius1973} goes beyond merely converting a nonlocal Lagrangian into infinite order. That conversion serves only to leverage the variational principle of mechanics along with the other mathematical tools that it entails. His ultimate goal is to derive the functional versions of all of these. This is how he proceeds with the formal sum of the series in (\ref{ei4}). Applying the chain rule to (\ref{ei3}) and using (\ref{ei2}), we have that
\beq \label{ei5}
\frac{\partial L}{\partial q^{(l)}} = \int_\RR \D\tau\,\frac{\delta \mathcal{L}\left(T_t q\right)}{\delta q(t+\tau)}\,\frac{\tau^l}{l!} \,.
\eeq 
Substituting this back into (\ref{ei4}) gives
$$ \mathcal{E}\left(q^{(k)}(t)\right) \equiv \int_\RR \D\tau\,\sum_{l=0}^\infty (-1)^l \,\frac{\tau^l}{l!} \,\frac{\partial^l\;}{\partial t^l} \left(\frac{\delta \mathcal{L} \left(T_t q\right) }{\delta q(t+\tau)} \right)\,.$$
Recognizing the formal Taylor sum,
$$ \sum_{l=0}^\infty \frac{(-\tau)^l}{l!} \,\frac{\partial^l\;}{\partial t^l} \left(\frac{\delta \mathcal{L} \left(T_t q \right) }{\delta q(t+\tau)} \right) = \frac{\delta \mathcal{L} \left(T_{t-\tau} q \right) }{\delta q(t-\tau+\tau)} = \frac{\delta \mathcal{L} \left(T_{t-\tau} q \right) }{\delta q(t)} \, $$
and changing integration variable to $\,\sigma = t - \tau\,$ yields the compact functional form
\beq \label{ei6}
\mathcal{E}\left(T_t q\right) \equiv \int_\RR \D\sigma\,\lambda(q, \sigma,t) \,, \qquad {\rm with} \qquad \lambda(q, \sigma,t):=
\frac{\delta \mathcal{L} \left(T_\sigma q\right) }{\delta q(t)} \,,
\eeq 
which is the EL equation written directly in terms of the nonlocal Lagrangian $\mathcal{L}$. Notice that, due to the relation (\ref{ei2}), $T_t q$ is ``equivalent'' to $\,q^{(k)}(t)\,,\;\, k=0,\,1,\,2\, \ldots\,$.

Similarly, the momenta (\ref{ei4c}) can be written, via (\ref{ei5}), as
\beq \label{ei9}
p_j\left(q^{(k)}(t)\right) = \int_\RR \D\tau\,\sum_{l=0}^{\infty} (-1)^l \,\mathbf{D}^l \left(\frac{\delta \mathcal{L} \left(T_t q\right) }{\delta q(t+\tau)} \right)\,\frac{\tau^{j+l+1}}{(j+l+1)!}  \,.
\eeq
Using the identity 
$$ \frac{\tau^{j+l+1}}{(j+l+1)!} = \frac1{j! \,l!}\,\int_0^\tau \D\tau^\prime \tau^{\prime j} (\tau - \tau^\prime)^l  \, $$
and including (\ref{ei6}), one finds
\begin{eqnarray*}
p_j\left(T_t q\right) & =& \int_\RR \D\tau\,\int_0^\tau \D\tau^\prime \frac{\tau^{\prime j}}{j!}\,\sum_{l=0}^{\infty} \frac{(\tau^\prime - \tau)^l}{l!}
\, \mathbf{D}^l \lambda(q,t,t+\tau)\, \\[1.5ex]
  & = & \int_\RR \D\tau\,\int_0^\tau \D\tau^\prime \,\frac{\tau^{\prime j}}{j!}\,\lambda\left(q,t+ \tau^\prime-\tau,t+\tau^\prime \right) \,.
\end{eqnarray*} 
Equivalently, this double integral can be written with step functions as
\beq \label{ei9a}
p_j\left(T_t q\right) = \int_{\RR^2} \D\tau\,\D\tau^\prime\,\left[\theta(\tau^\prime) - \theta(\tau^\prime-\tau) \right]\, \frac{\tau^{\prime j}}{j!}\,\lambda\left(q,t+ \tau^\prime-\tau,t+\tau^\prime \right)\,.
\eeq

Substituting the latter into  the series term on the left hand side of (\ref{ei8c}) and changing integration variables to $\sigma=\tau^\prime$ and $\rho =\tau^\prime-\tau\,$ yields
$$ \sum_{j=0}^\infty p_{j}\left(T_t q\right)\,\delta q^{(j)}(t) = \int_{\RR^2} \D\rho\,\D\sigma\,\left[\theta(\sigma) - \theta(\rho) \right]\, \lambda\left(q,t+ \rho,t+\sigma\right)\, \delta q(t+\sigma) \, $$
and, taking $t=0\,$, 
\beq  \label{ei10z}
\sum_{j=0}^\infty p_{j}\left(q\right)\,\delta q^{(j)} = \int_{\RR^2} \D\rho\,\D\sigma\,\left[\theta(\sigma) - \theta(\rho) \right]\, \lambda\left(q,\rho,\sigma\right)\, \delta q(\sigma) \,.
\eeq
Likewise, the energy (\ref{ei8b}) becomes
$$ E\left(T_t q\right) =\int_{\RR^2} \D\rho\,\D\sigma\,\left[\theta(\sigma) - \theta(\rho) \right]\, \lambda\left(q,t+ \rho,t+\sigma\right)  \, \dot q(t+\sigma)  - \mathcal{L} \left(T_t q \right) $$
and, taking $t=0\,$, 
\beq  \label{ei10a}
E\left(q\right) =\int_{\RR^2} \D\rho\,\D\sigma\,\left[\theta(\sigma) - \theta(\rho) \right]\, \lambda\left(q,\rho,\sigma\right)\,
\dot q(\sigma)  - \mathcal{L} \left(q \right)\,. 
\eeq
Combining then equations (\ref{ei10z}-\ref{ei10a}) in (\ref{ei8c}), we obtain the functional form of the conserved charge (\ref{ei8c}):
\beq  \label{ei10b}
Q\left(q \right) :=  \int_{\RR^2} \D\rho\,\D\sigma\,\left[\theta(\sigma) - \theta(\rho) \right]\, \lambda\left(q,\rho,\sigma\right)\,\left[ \delta q(\sigma) - \dot q(\sigma)\, \delta T \right] - \mathcal{L} \left(q \right)\, \delta T \,.
\eeq
A similar construction yields a functional expression for the presymplectic form (\ref{ei11a}) as the differential of (\ref{ei10z}). 

\subsection{An inconsistency \label{Paradox}}
The drawback of the functional equations (\ref{ei6}), (\ref{ei9a}), (\ref{ei10a}), and (\ref{ei10b}) is that they rest on heuristic reasoning. Consequently, certain inconsistencies can emerge ---most notably those associated with the endpoint conditions at $t_1$ and $t_2$, as we discuss below.

Consider a nonlocal action defined by the nonlocal Lagrangian $\,\mathcal{L}\left(T_t q \right)$ and its infinite-order counterpart ${L}\left(q^{(j)}(t) \right)\,$. Their corresponding nonlocal action integrals are
$$ S_{NL}(q,t_1,t_2) := \int_{t_1}^{t_2} \D t\,\mathcal{L}\left(T_t q \right) \qquad {\rm and} \qquad S_\infty(q,t_1,t_2) := \int_{t_1}^{t_2} \D t\,{L}\left(q^{(j)}(t) \right)   \,,$$
where $S_\infty$ is pseudo-local (expressed in terms of infinitely many derivatives at each instant) and $S_{NL}$ is fully nonlocal (depending on the entire trajectory segment).

The EL equation derived from the pseudo-local action $S_\infty$ is (\ref{ei4}) that after a formal addition yields (\ref{ei6}):
\begin{equation}  \label{Par1}
\mathcal{E}_\infty(q,t) \equiv \int_\RR \D\sigma \frac{\delta \mathcal{L}\left(T_\sigma q \right)}{\delta q(t)} = 0 \,, \qquad t \in [t_1,t_2] \,.
\end{equation}
 
In contrast, varying the fully nonlocal action $S_{NL}$ gives  
$$ \delta S_{NL} \equiv  \int_{t_1}^{t_2} \D\tau \int_\RR \D\sigma \,\frac{\delta \mathcal{L}\left(T_\tau q \right)}{\delta q(\sigma)}\,\delta q(\sigma)   
\equiv \int_\RR \D\sigma\,\delta q(\sigma)\, \int_{t_1}^{t_2} \D\tau \frac{\delta \mathcal{L}\left(T_\tau q \right)}{\delta q(\sigma)} \,, $$
so its EL equation is
\begin{equation}  \label{Par2}
\mathcal{E}_{NL}(q,\sigma) \equiv \int_{t_1}^{t_2} \D\tau \frac{\delta \mathcal{L}\left(T_\tau q \right)}{\delta q(\sigma)} = 0 \,, \qquad\forall \sigma \in \RR \,.
\end{equation}
Clearly, $\mathcal{E}_\infty$ and $\mathcal{E}_{NL}$ differ in their domains of integration, reflecting an inconsistency when the pseudo-local and fully nonlocal formulations are compared.

\section{Nonlocal Lagrangians \label{S5}}
Nonlocal Lagrangians depend functionally on the trajectory, whether on the entire trajectory, on a finite segment of it, or on the values of $q$ and some of its derivatives at a finite number of instants of time. Here, we develop a rigorous framework for nonlocal Lagrangians without rewriting them as infinite-order derivative expansions, thereby circumventing the inconsistencies discussed in Section~\ref{S4} associated to the use of formal Taylor series expansions such as~(\ref{ei2}). Within this framework, we will establish, on firm mathematical grounds, precisely the same results derived in Section~\ref{S4} on a heuristic basis, namely the EL equation, Noether's theorem, and the Hamiltonian formalism. The heuristic arguments previously employed will be used here only as a motivation for the fundamental principles; once these have been established, we proceed by mechanical application of mathematical rules.

\subsection{The variational principle \label{S5.1}}
A nonlocal Lagrangian is a functional $\mathcal{L}$ defined on the space of trajectories, referred to as the \emph{kinematic space} and generically denoted by $\mathcal{K}$. Moreover, this space is taken to be $\mathcal{C}^k(\mathbb{R},\mathbb{R}^m)$ for some integer $k$, with $m$ the number of degrees of freedom. Here, for simplicity, we restrict ourselves to the case of a single degree of freedom.

In the infinite-order representation, the action integral typically has the form
\[
\int_{t_1}^{t_2}\D t\, L\bigl(q(t),\dot q(t),\dots,q^{(n)}(t),\dots\bigr),
\]
where the argument of the Lagrangian contains information about the trajectory in the vicinity of the instant $t$. Formally, according to the Taylor expansion~(\ref{ei2}), the same information is encoded by the translated trajectory $T_t q(\tau)=q(t+\tau)$. Thus, one might initially consider an action integral of the form $\int_{t_1}^{t_2}\D t\,\mathcal{L}(T_t q)$ for fixed endpoints $t_1,\,t_2$. However, to avoid the mathematical inconsistencies highlighted in Section~\ref{Paradox}, it is preferable to integrate over the entire real line, although this introduces the potential issue of divergent integrals. To address this issue (circumvent this \ldots), instead of using a single action integral, we introduce a two-parameter family of action integrals:
\begin{equation}\label{nl1} 
S(q,t_1,t_2)=\int_{t_1}^{t_2}\D t\,\mathcal{L}(T_t q),\qquad t_1,t_2\in\mathbb{R}\,,
\end{equation}
and then propose the following \emph{nonlocal variational principle}:
\begin{equation}\label{nl2}
\lim_{-t_1,t_2\rightarrow\infty}\|\delta S(q,t_1,t_2)\|=0,\qquad\text{for all variations }\delta q(t)\text{ with bounded support.}
\end{equation}

The norm in the last equation needs to be explained. $S(q,t_1,t_2)$ being a functional on the Banach space $\,\mathcal{C}(\left[t_1,t_2\right], \RR)\,$, $\,\delta S(q,t_1,t_2)\,$ belongs to the Banach space of differential 1-forms on it.  Then the symbol $\|\delta S(q,t_1,t_2)\|$ means the norm in the Banach space of differential 1-forms \cite{CartanH}, namely:
$$ 
\|\delta S(q,t_1,t_2)\|=\sup \left\{\left|\frac{\delta S(q,t_1,t_2)}{\delta q(\sigma)}\right|, \, t_1 \leq \sigma \leq t_2\,\right\}
$$
and the variational principle~(\ref{nl2}) explicitly means
\begin{equation}\label{nl3}
\lim_{-t_1,t_2\rightarrow\infty} \sup \left\{\left|\frac{\delta S(q,t_1,t_2)}{\delta q(\sigma)}\right|, \, t_1 \leq \sigma \leq t_2\,\right\} = 0 \,,
\end{equation}
which is clearly equivalent to 
$$
\lim_{-t_1,t_2\rightarrow\infty} \frac{\delta S(q,t_1,t_2)}{\delta q(\tau)}= 0 \,, 
\qquad  \, \forall \tau \in \RR\,.
$$
Consequently, we arrive at the EL equation in integral form:
\begin{equation}\label{nl4}
\mathcal{E}(q,\tau)\equiv\int_{\mathbb{R}}\D t\,\lambda(q,t,\tau)=0,\qquad\text{where}\quad\lambda(q,t,\tau)=\frac{\delta\mathcal{L}(T_t q)}{\delta q(\tau)},\quad\text{for all }\tau\in\mathbb{R}.
\end{equation}
Equation~(\ref{nl4}) precisely recovers the EL equation (\ref{ei6}) derived in Section~\ref{S4} on a heuristic basis, but now established on rigorous mathematical grounds.

Notice that
\beq \label{nl4a} 
\lambda\left(T_\sigma q,t,\tau\right) = \frac{\delta \Lcal\left(T_t T_\sigma q\right)}{\delta T_\sigma q(\tau)} = \frac{\delta \Lcal \left(T_{\sigma+t} q\right)}{\delta q(\sigma+\tau)} =  \lambda(q,\sigma+t,\sigma+\tau)  \,,
\eeq
and therefore from (\ref{nl4}) it follows that
$$ \mathcal{E} (T_\sigma q, \tau) \equiv \int_\RR \D t  \, \lambda(T_\sigma q,t ,\tau)
 \equiv \int_\RR \D t  \, \lambda(q,t +\sigma,\tau+\sigma) \equiv  \mathcal{E} (q, \tau+\sigma)  \,. $$
Hence, the EL equation is invariant under time translations:
\beq \label{nl4b}  
\mathcal{E} (T_\sigma q, \tau) = 0\,, \quad \forall\,\tau \in \RR \qquad \mbox{if, and only if} \qquad 
  \mathcal{E} (q, \tau+\sigma) = 0 \,, \quad \forall\,\sigma \in \RR \,.
\eeq

The phase space $\mathcal{D}\,$ consists of all trajectories  $\,q(\tau) \in \mathcal{K}\,$ satisfying the EL equation, which is of functional type. For such nonlocal equations, there are not general existence and uniqueness theorems as in the regular local Lagrangian case. Therefore, equation (\ref{nl4}) cannot be taken as an evolution law for an initial state, but rather must be interpreted as an implicit equation ---or, more precisely, a one-parameter family of constraints--- that defines $\mathcal{D}\,$ as a submanifold of $\mathcal{K}\,$.

Including what has been discussed in Section \ref{PhSp}, it is obvious that this dynamic space $\mathcal{D}$ is the phase space of the nonlocal system. Then the problem is to find a minimal parameterization that unambiguously characterizes each trajectory; this task must be addressed specifically for each case, depending on the type of the EL equation, as we do in Section \ref{S6} for a number of examples.


Time evolution in the kinematic space is given by (\ref{ei2}):
$$ T_t q(\tau) := q(t+\tau) \,, $$
whose infinitesimal generator is
\beq \label{nl9y} 
 \mathbf{D} f(q,t) := \left[\frac{\D f(T_\tau q,t+\tau)}{\D \tau} \right]_{\tau=0} \,,
\eeq
for any function $f$ defined on $\mathcal{K}\,$. In particular, one has $\,\mathbf{D} q(\tau) = \dot q(\tau)\,$.

Combining (\ref{nl9y}) and (\ref{nl4b}), we find
$$ \mathbf{D}\mathcal{E} (q,\tau) \equiv \partial_\tau \mathcal{E}(q,\tau)\,.$$
Thus, the constraints $\,\mathcal{E}(q,\tau) = 0\,,\;\,\tau\in \RR\,,$ are stable under the time evolution, or equivalently, the vector field  $\mathbf{D}$ is tangent to the phase space $\mathcal{D}\,$.

\subsection{Noether's theorem for nonlocal Lagrangians \label{S5.2}}
For finite-order Lagrangians, the canonical momenta ---which the Legendre transformation is based on--- arise naturally as the prefactors of $\delta q^{(j)}$ in the boundary terms obtained through integration by parts when the variational principle is developed. A similar procedure applies to infinite-order Lagrangians, though the heuristic nature of that derivation makes its validity less stringent. In both finite and infinite-order cases, these canonical momenta also appear as the prefactors of $\delta q^{(j)}$ within the conserved quantities derived from Noether's theorem.

However, for nonlocal Lagrangians, no integration by parts occurs when implementing the variational principle, thus leaving no immediate method to define canonical momenta or generalize the Legendre transformation. We therefore pursue a generalization of Noether's theorem to the nonlocal setting and aim to guess the canonical momenta from the resulting conserved quantities.

Consider an infinitesimal transformation of the form
\begin{equation}  \label{nl6}
t^\prime(t) = t + \delta T(t) \,, \qquad \quad q^{\prime}(t^\prime) = q(t) + \delta q(t) \,
\end{equation}
and apply it as a change of variables in the action integral~(\ref{nl1}).  For any given $t_1$, $t_2$, one obtains the identity
\begin{equation}  \label{nl7}
 \int_{t^\prime_1}^{t^\prime_2} \mathcal{L}^\prime\left(T_{t^\prime}q^\prime\right)\,\D t^\prime \equiv \int_{t_1}^{t_2} \mathcal{L}\left(T_t q\right)\,\D t 
\end{equation}
from which it follows that the induced transformation of the Lagrangian is given by
\[
\mathcal{L}^\prime\left(T_{t^\prime}q^\prime\right) = \mathcal{L}\left(T_{t(t^\prime)} q(q^\prime) \right)\,\frac{\D t}{\D t^\prime}\,.
\]  

The transformation (\ref{nl6}) is called a symmetry transformation of the Lagrangian if
\begin{equation}  \label{nl8}
 \mathcal{L}^\prime\left(T_{t^\prime}q^\prime\right) \equiv \mathcal{L}\left(T_{t^\prime} q^\prime \right)\,.
\end{equation}
In such a case, by renaming the dummy integration variable in~(\ref{nl7}) and neglecting second-order infinitesimals, we obtain:
\begin{eqnarray*}  
 0 &\equiv & \int_{t^\prime_1}^{t^\prime_2} \mathcal{L}\left(T_\tau q^\prime\right)\,\D \tau -\int_{t_1}^{t_2} \mathcal{L}\left(T_\tau q \right)\,\D \tau  \\[1ex]   
  &\equiv & \mathcal{L}\left(T_{t_2}q \right)\,\delta T_2 - \mathcal{L}\left(T_{t_1}q \right)\,\delta T_1 + \int_{t_1}^{t_2} \D \tau \, \left[\mathcal{L}\left(T_\tau q^\prime \right) - \mathcal{L}\left(T_\tau q \right) \right]  \\[1ex]   
  &\equiv &  \int_{t_1}^{t_2} \D \tau \,\left\{  \int_\RR \D\sigma\,\frac{\delta \mathcal{L}\left(T_\tau q \right)}{\delta q(\sigma)}\,\tilde\delta q(\sigma) + \partial_\tau \left[\mathcal{L}\left(T_{\tau}q\right)\,\delta T(\tau)\right] \right\}\,,
\end{eqnarray*}
where 
\begin{equation}  \label{tildedelta}
\,\tilde\delta q(\sigma) := q^\prime (\sigma) - q(\sigma) \equiv \delta q(\sigma) - \dot q(\sigma)\,\delta T(\sigma) \,.
\end{equation}
Adding (\ref{nl4}) into the expression above, we can continue the string of equalities as
\begin{eqnarray}  \label{nl9}
 0 &\equiv & \int_{t_1}^{t_2}\hspace*{-.5em} \D \tau\,\left\{\mathcal{E}(q,\tau)\,\tilde\delta q(\tau)  + \int_\RR \D\sigma\,\left[- \lambda(q,\sigma,\tau)\,\tilde\delta q(\tau) +  
\lambda(q,\tau,\sigma)\,\tilde\delta q(\sigma) \right]+ \right.\nonumber\\[1.5ex] \label{nl9aa}
 & & \hspace*{3em}\left. \partial_\tau \left[\mathcal{L}\left(T_{\tau}q\right)\,\delta T(\tau)\right] \;\right\} \,. 
\end{eqnarray}
The two integrals over $\sigma$ in the right-hand side can be simplified by the change of variables $\sigma = \tau - \rho$ in the first and $\sigma = \tau + \rho$ in the second. This yields the combined expression:
\beq \label{nl9z}
\int_\RR \D\rho\,\left[\lambda(q,\tau,\tau+\rho)\,\tilde\delta q(\tau+\rho) - \lambda(q,\tau-\rho,\tau)\,\tilde\delta q(\tau)\right] \,.
\end{equation}
Now, observe the identity:
\begin{eqnarray*}
\hspace*{-3em}\lefteqn{\lambda(q,\tau,\tau+\rho)\,\tilde\delta q(\tau+\rho) -\lambda(q,\tau-\rho,\tau)\,\tilde\delta q(\tau) \equiv } \\[1ex]
 \qquad &  \qquad \quad \equiv &  \int_0^1\D\eta \,\frac{\partial\;}{\partial\eta}\left[\lambda(q,\tau+(\eta-1)\rho,\tau+\eta\rho)\,\tilde\delta q(\tau+\eta\rho) \right] \\[2ex]
 &  \qquad \quad \equiv & \rho\, \frac{\partial\;}{\partial \tau}\,\int_0^1\lambda(q,\tau+(\eta-1)\rho,\tau+\eta\rho)\,\tilde\delta q(\tau+\eta\rho)\,\D\eta \\[2ex]
 & \qquad \quad \equiv & {\partial_\tau}\,\int_0^{\rho} \lambda(q,\tau+\xi-\rho,\tau+\xi)\,\tilde\delta q(\tau+\xi)\,\D\xi  \,,
\end{eqnarray*}
where we have used the change of variables $\xi = \eta\rho$. Now using (\ref{nl4a}), the integral in~(\ref{nl9z}) can be written as
\beq \label{nl9a}
\int_\RR \D\rho\,\left[\lambda(q,\tau,\tau+\rho)\, \tilde\delta q(\tau+\rho) - \lambda(q,\tau-\rho,\tau)\, \tilde\delta q(\tau)\right] \equiv 
\partial_\tau \Pi\left(T_\tau q\right) \,,
\end{equation}
where we define
\begin{eqnarray}
 \Pi(q) &:=& \int_\RR \D\rho\,\int_0^{\rho}\hspace*{-.3em}\D\xi\, \lambda(q,\xi-\rho,\xi)\, \tilde\delta q(\xi)  
\nonumber  \\[2ex]   
\label{nl9b}
  & \equiv & \int_{\RR^2} \D\rho\,\D\xi\, \left[\theta(\xi) - \theta(\xi-\rho)\right] \,\lambda(q,\xi-\rho,\xi)\, \tilde\delta q(\xi)\,. 
\end{eqnarray}
Changing variables to \(\sigma = \xi - \rho\), we obtain an equivalent expression:
\begin{equation} \label{nl9c}
 \Pi(q) \equiv \int_\RR \D\xi\,p(q,\xi)\, \tilde\delta q(\xi) \,,\quad {\rm where} \quad 
p(q,\xi) := \int_{\RR} \D\sigma\,\left[ \theta(\xi)-\theta(\sigma) \right]\,\lambda(q,\sigma,\xi) \,.
\end{equation}
Substituting~(\ref{nl9a}) into~(\ref{nl9}), we obtain
$$  \int_{t_1}^{t_2} \D \tau \,\left\{\mathcal{E}(q,\tau)\, \tilde\delta q(\tau) +\frac{\partial }{\partial \tau} \left[\mathcal{L}\left(T_{\tau}q\right)\,\delta T(\tau) + \Pi(T_\tau q) \right]\right\} \equiv 0 \,,  $$
for arbitrary $t_1$ and $t_2$. Therefore, we arrive at the identity
\begin{equation}  \label{nl10}
\mathcal{E}(q,\tau)\, \tilde\delta q(\tau) +\frac{\partial \;}{\partial \tau} \left\{- E\left(T_{\tau}q\right)\,\delta T(\tau) + 
\int_\RR \D\xi\,p(T_\tau q,\xi)\, \delta T_\tau q(\xi) \right\} \equiv 0 \,,  
\end{equation}
where expressions~(\ref{tildedelta}) and~(\ref{nl9c}) have been used. Here, the quantity
\begin{equation}  \label{nl14}
 E(q) :=  \int_\RR \D\xi\,p(q,\xi)\,\dot q(\xi) - \mathcal{L}(q)
\end{equation}
plays the role of the \emph{energy}. 

The equality~(\ref{nl10}) is the \emph{Noether identity} for nonlocal Lagrangians. It is an \emph{off-shell} statement, meaning it holds for any kinematic trajectoriy \( q(\tau) \), regardless of whether it satisfies the equations of motion. For dynamical trajectories, one has \(\mathcal{E} = 0\), and the Noether identity~(\ref{nl10}) reduces to the conservation law:
$$ \mathbf{D} Q = 0 \,, \qquad {\rm where} \qquad Q(q) := \int_\RR \D\xi\,p(q,\xi)\,\delta q(\xi) - E\left(q\right)\,\delta T(\tau) \,  $$
and \(\mathbf{D}\) is the generator of time evolution introduced in~(\ref{nl9y}):
$$ \mathbf{D} f(q,t) := \left[\frac{\D f(T_\tau q,t+\tau)}{\D \tau} \right]_{\tau=0} \,. $$
We have thus established the extension of Noether's theorem to nonlocal Lagrangians: any symmetry transformation that preserves the form of the nonlocal Lagrangian gives rise to a conserved quantity along dynamical trajectories, that is:

\begin{theorem}
Let the infinitesimal transformation~\eqref{nl6} be a symmetry of the nonlocal action, in the sense that it preserves the Lagrangian. Then, the associated conserved quantity for dynamical trajectories is
\begin{equation}  \label{nl11}
Q(q,\tau) :=  \int_\RR \D\xi\,p(q,\xi)\,\delta q(\xi)- E\left(q\right)\,\delta T(\tau)\,, 
\end{equation}
where \(p(q,\xi)\) is defined in~\eqref{nl9c} and \(E(q)\) is the energy functional given in~\eqref{nl14}.
\end{theorem}
Notice the structural similarity of this expression with equation~(\ref{en8b}) derived in the context of higher-order Lagrangians.

\paragraph{The energy:} Infinitesimal time translations are defined by $\,t^\prime = t + \varepsilon\,$ and $\,q^\prime(t^\prime) = q(t)\,$, which implies that the transformation~(\ref{nl6}) becomes
$\, \delta t(\tau) = \varepsilon\,,\,\, \delta q(\tau)= 0 \,$.
If the Lagrangian does not depend explicitly on $t\,$, it is invariant under time translations, and the associated conserved charge (\ref{nl11}) becomes $\,Q = - \varepsilon\, E \,.$ Substituting this into the Noether identity~(\ref{nl10}) evaluated at \( \tau = 0 \), we obtain the identity:
\begin{equation}  \label{nl12z}
\mathbf{D} E(q) + \dot q(0)\,\mathcal{E}(q,0) \equiv 0.
\end{equation}
Therefore, the energy is preserved on dynamical trajectories.

\subsection{Hamiltonian formalism \label{SS.5.2}}
We proceed in the same spirit as in Sections~\ref{S2.2}, \ref{S3.2}, and~\ref{S4.2}, where differential forms on the phase space were a fundamental tool. In the nonlocal case, the relevant space is the kinematic space which has infinitely many dimensions. Now $\mathcal{K}$ is  $\mathcal{C}^k(\RR,\RR^m)\,$, for some positive integer $k$, and trajectories $\,q \in \mathcal{K}\,$ may be unbounded. Thus it may happen that $\,\sup\,\{\|q^{(j)}(t)\|,\,t\in \RR, 0\leq j\leq k\}\,$ does not exist. Consequently, $\mathcal{K}$ is not a Banach space and the differential calculus on it \cite{CartanH,Choquet} cannot be applied to our case.

However, the countable family of seminorms on $\,\mathcal{C}^k(\RR,\RR^m)\,$
$$ P_n(q) = \sup \{\|q^{(j)}(t)\|,\,|t| \leq n , 0\leq j\leq k\} \,, \qquad n \in \NN \,, $$
define $\mathcal{K}$ as a Fr\'echet space \cite{Choquet, Apostol}. We thus have differential calculus on such a space, as well as a way of defining differential forms on $\mathcal{K}\,$, which provides a sound mathematical basis for what follows\footnote{A great number of results proved in \cite{CartanH} concerning differential calculus on Banach spaces can be extended to Fr\'echet spaces, particularly Poincar\'e Lemma on closed differential forms.}.

We will not adopt here the technical notation in the cited references because it is too cumbersome and rather unfamiliar to the audience we are aiming to, namely theoretical physicists. Instead we shall adopt a more naive but understandable and practical notation, which basically replaces $\displaystyle{\sum_{i=1}^n \ldots}\,$ and $\,\D q \,$, in the differential forms for a finite number of dimensions, with $\,\displaystyle{\int_\RR \D\lambda \, \ldots}\,$ and $\delta q(\lambda)\,$, respectively. Here, $\delta$ denotes the differential acting on functions on the Fr\'echet space, to avoid confusion with Leibniz's notation in the integrals.

We begin by introducing the (pre)symplectic form
\begin{equation}  \label{nl15}
\omega(q) :=  \int_\RR \D\xi\,\delta p(q,\xi) \wedge \delta q(\xi)  \,,
\end{equation}
that is, a closed differential 2-form on the kinematic space\footnote{We use the symbol $\delta$ instead of $\D$ to denote the differential acting on functions on the Banach space $\mathcal{K}$, to avoid confusion with Leibniz's notation in the integrals.}, where the momenta $p(q,\xi)$ are given by the prefactors of $\,\delta q(\xi)\,$ in the conserved quantity (\ref{nl11}). The momenta can be thought as a map defined on the kinematic space $\mathcal{K}$:
$$\, q \longrightarrow p\,, \quad\text{where $\,p(q)\,$ is the function} \,\,\, p(q)_{(\xi)} := p(q,\xi)$$
defined by the expression (\ref{nl9c}). Next, we take the energy functional defined in~(\ref{nl14}):
$$\,E(q):=  \int_\RR \D\sigma\,p(q,\sigma)\,\dot q(\sigma) - \mathcal{L}(q) \, $$
and introduce the 2-form $\, \beta := \omega - \delta E \wedge \delta t \,$ on the extended kinematic space $\,\mathcal{K}^\prime := \mathcal{K} \times \RR \,$.

As we did in the local cases, we compute the inner product of the vector field \( \mathbf{D} \) with the 2-form \( \beta \). Using~(\ref{nl12z}), we write: 
\[
\,i_\mathbf{D} \beta = i_\mathbf{D} \omega + \delta E + \mathcal{E}(q,0)\,\dot q(0)\,\delta t\,.
\]
We also compute:
\begin{eqnarray*}
 i_\mathbf{D} \omega &\equiv& \int_\RR \D\sigma\,\left\{\mathbf{D}p(q,\sigma)\,\delta q(\sigma) - \mathbf{D} q(\sigma)\, \delta p(q,\sigma)\right\} \,, \\[1.5ex]
 \delta E &\equiv& \int_\RR \D\sigma\,\left\{p(q,\sigma)\,\delta\dot q(\sigma) + \dot q(\sigma)\, \delta p(q,\sigma) - \lambda(q,0,\sigma)\,\delta q(\sigma) \right\}\,,
\end{eqnarray*}
where $\;\mathbf{D} q(\sigma) \equiv  \partial_\sigma q(\sigma) \,$. Combining the two expressions yields:
$$ i_\mathbf{D} \beta \equiv \int_\RR \D\sigma\,\left\{\left[\mathbf{D}p(q,\sigma) - \lambda(q,0,\sigma)\right]\,\delta q(\sigma) + p(q,\sigma)\,\delta\dot q(\sigma) \right\} + \mathcal{E}(q,0)\,\dot q(0)\,\delta t\,,$$
which, after integrating by parts, gives:
\begin{equation}  \label{nl15z}
i_\mathbf{D} \beta \equiv \int_\RR \D\sigma\,\left\{\mathbf{D}p(q,\sigma) - \partial_\sigma p(q,\sigma)-  \lambda(q,0,\sigma) \right\} \,\delta q(\sigma) + \mathcal{E}(q,0)\,\dot q(0)\,\delta t
\end{equation}
Now, using (\ref{nl4a}), one verifies that $\,\mathbf{D} \lambda(q,\xi,\sigma) \equiv \left[\partial_\xi + \partial_\sigma\right]\,\lambda(q,\xi,\sigma) \,,$ which, together with (\ref{nl9c}) and after some algebra, leads to
$$ \mathbf{D}p(q,\sigma) \equiv \partial_\sigma p(q,\sigma) + \lambda(q,0,\sigma) - \delta(\sigma)\,\mathcal{E}(q,0) \,, $$
where we have used that \( \lambda(q,\pm\infty,\sigma) = 0 \), because \( \mathcal{E}(q,\sigma) \) must be integrable. Substituting the latter into~(\ref{nl15z}), we finally obtain
\begin{equation}  \label{nl15y}
i_\mathbf{D} \beta \equiv -\mathcal{E}(q,0)\,\left[\delta q(0)  -\dot q(0)\,\delta t\right]  \,.
\end{equation}
This identity is the nonlocal counterpart of similar expressions previously derived for local systems, namely~(\ref{en9b}) in the finite-order case and~(\ref{ei9b}) in the infinite-order case.


\subsection{The presymplectic form and the energy \label{SS.5.3}}
Next, we work out the expressions~(\ref{nl14}) and~(\ref{nl15}) in more detail, as the energy and the presymplectic form will be useful in the examples developed in Section~\ref{S6}. 

Combining the definition of the presymplectic form~(\ref{nl15}) with the expression for the momenta in~(\ref{nl9c}), we obtain:
$$\omega = \int_\RR \D\xi\, \int_\RR \D\sigma\, \left[\theta(\xi)-\theta(\sigma)\right]\, \delta \lambda(q,\sigma,\xi) \wedge \delta q(\xi)   \,  $$
which, upon inserting~(\ref{nl4}), becomes:
$$ \omega = \int_\RR \D\xi\,\int_\RR \D\tau\, \int_\RR \D\sigma\, \left[\theta(\xi)-\theta(\sigma)\right]\, \frac{\delta^2 \Lcal \left(T_\sigma q\right)}{\delta q(\xi) \, \delta q(\tau)}\,\delta q(\tau) \wedge \delta q(\xi)  \,$$
and, using the skewsymmetry of the wedge product, we get
\begin{equation}  \label{nl18}
\omega = \int_\RR \D\xi\,\int_\RR \D\tau\, \frac12\, \left[\theta(\xi)-\theta(\tau)\right]\,\Omega(q,\xi,\tau) \,\delta q(\tau) \wedge \delta q(\xi)  \,,
\end{equation}
where 
\begin{equation}  \label{nl19a}
 \Omega(q,\xi,\tau) :=\int_\RR \D\sigma\,\frac{\delta^2 \Lcal \left(T_\sigma q\right)}{\delta q(\xi) \, \delta q(\tau)}  \,. 
\end{equation}
Note that the factor \( \theta(\xi) - \theta(\tau) \) vanishes unless either  $\xi<0$ and $\tau\geq 0$, or $\tau<0$ and $\xi\geq 0$, in which case it is \(-1\) and \(1\), respectively. Therefore (\ref{nl18}) becomes 
\begin{equation}  \label{nl19b}
 \omega = \int_0^\infty \D\xi\,\int_{-\infty}^0 \D\tau\, \Omega(q,\xi,\tau) \,\delta q(\tau) \wedge \delta q(\xi)  \,. 
\end{equation}
We now introduce the functions defined on the positive real half-line:
\begin{equation}  \label{nl18a}
Q(\tau):= q(\tau)\,, \qquad  U(\tau):= q(-\tau) \,, \qquad 0 \leq \tau \,,
\end{equation}
which allow us to reconstruct the full trajectory via
\begin{equation}  \label{nl18z}
 q(\tau) = \left\{ \begin{array}{ll}
                      Q(\tau) \qquad &\,,\tau\geq 0 \\
											U(|\tau|) \qquad &\,, \tau\leq 0\,.
											\end{array}    \right.  
\end{equation}
Using this decomposition, we can rewrite the presymplectic form~(\ref{nl18}) as 
\begin{equation}  \label{nl18b}
\omega = \int_0^\infty \D\xi\,\int_0^\infty \D\tau\, \Omega(Q,U,\xi,-\tau) \,\delta U(\tau) \wedge \delta Q(\xi)  \,.
\end{equation}

\paragraph{The energy.} We may proceed similarly with the definition~(\ref{nl14}) to obtain the following expression:
\begin{equation}  \label{nl18c}
 E(q) = \int_\RR \D\xi\, \int_\RR \D\sigma\, \left[\theta(\xi)-\theta(\sigma)\right]\, \lambda(q,\sigma,\xi) \,\dot q(\xi) - \Lcal(q) \,.
\end{equation}
Taking into account the support properties of the factor $\,\theta(\xi)-\theta(\tau)\,$, this can be rewritten as:
\begin{equation}  \label{nl20}
 E(q) = \int_0^\infty \D\xi\, \int_0^\infty \D\sigma\, \left[\lambda(q,-\sigma,\xi) \,\dot q(\xi) - \lambda(q,\xi,-\sigma) \,\dot q(-\sigma) \right] - \Lcal(q) \,,
\end{equation}
which is the appropriate form for expressing the Hamiltonian in terms of the half-line variables (\ref{nl18a}),  \( Q \) and \( U \).

The 2-form \( \omega \) is closed, as it is apparent in (\ref{nl15}), and in order for it to be a symplectic form, it must define a bijective map
$$ \mathbf{F}  \in T\mathcal{K} \longrightarrow i_\mathbf{F} \omega  \in \Lambda^1 \mathcal{K}\,, $$
as discussed in~\cite{Marsden}. If the map is merely injective, \( \omega \) is said to be \emph{weakly symplectic}. Since the expression~(\ref{nl18b}) contains only cross-terms of the form \( \delta U(\tau) \wedge \delta Q(\xi) \), it suffices to examine the invertibility of the linear operator
$$ u(\tau) \longrightarrow v(\xi) = \int_0^\infty \D\tau\, \Omega(Q,U,\xi,-\tau)\,u(\tau) \,, $$
acting on functions defined on the positive real half-line.

If this were the case, we would have endowed the kinematic space with a symplectic structure. However, the phase space is a submanifold defined implicitly by the EL equation~(\ref{nl4}):
$$ \mathcal{D} \stackrel{j}{\longrightarrow } \mathcal{K} \,, \qquad \mathcal{E}(q,\tau) = 0\,,\quad \forall \tau\in \RR \,. $$

The injection map \( j \) allows us to pull back the 2-form \( \omega \in \Lambda^2\mathcal{K} \) to the phase space, yielding \( j^*\omega \in \Lambda^2\mathcal{D} \). Thanks to the constraints, equation~(\ref{nl15y}) implies
$$ i_\mathbf{D}\beta \equiv i_\mathbf{D}\omega + \delta E = 0  \,,$$
which are precisely the Hamilton equations on the phase space.

What remains to be shown is that the pulled-back form \( j^* \omega \in \Lambda^2\mathcal{D} \) is symplectic ---that is, non-degenerate--- and to construct appropriate canonical coordinates. Although we have not a general answer to this issue, we address this question in the context of specific examples in the next section.

\section{Some examples \label{S6}}
Next, we implement the procedures presented in Section \ref{S5} to some illustrative examples: a nonlocal finite harmonic oscillator, the fully nonlocal Pais-Uhlenbeck model and a half-advanced half retarded oscillator, to show how the phase space, the presymplectic structure, the Hamiltonian, etc are determined. As commented in Section \ref{S5}, the EL equation is to be taken as an implicit equation defining the phase space although, in many instances the number of dimensions of $\mathcal{D}$ and a suitable coordinatization of it can be found without solving this implicit equation.

\subsection{Nonlocal (finite) Harmonic Oscillator \label{sec:nonlocal-oscillator} }
This example  consists of a \emph{nonlocal harmonic oscillator}, whose coordinate \(q(t)\)  is  tied to its entire temporal trajectory by the exponentially decaying kernel \(k(\sigma)=e^{-|\sigma|}\). Physically, we picture this phenomenon as a temporal spring: a disturbance at time $t$ feels a restoring pull from both its causal past and an anticipatory projection of its future\footnote{This time symmetry and the subsequent lack of causality are inherent to all nonlocal Lagrangian systems.}, with both influences fading exponentially as their temporal separation $|t-\tau|$ grows. This Lagrangian appears naturally upon integrating out one degree of freedom from a coupled system consisting of a harmonic oscillator interacting with a second coordinate subject to exponential decay. Consequently, it is referred to as a ``derived'' nonlocal Lagrangian in Ref.~\cite{EliezerWoodard1989}. Explicitly, it reads:
\begin{equation}\label{eq:LHO}
  \mathcal{L}\bigl(T_t q\bigr)=\frac12\dot q^{2}-\frac{\omega^{2}}{2}q^{2}
  +\frac g4\,q(t)\int_{\mathbb{R}} e^{-|t-\tau|}\,q(\tau)\,d\tau,
\end{equation}
where the coupling \(g\) sets the overall stiffness of this ``temporal spring,'' and contributions from distant times therefore become exponentially faint. Because the convolution in the Lagrangian must be finite for every time \(t\), the kinematic space is to be restricted to
\begin{equation}\label{eq:NLHOSC-K}
    \mathcal{K}
  =\bigl\{\, q\in \mathcal{C}^2(\mathbb{R})
     \ \big|\
      e^{-|\tau|}\,q(\tau) \,\mbox{ is summable in } \RR
   \bigr\}.
\end{equation}

The functional derivatives of the Lagrangian are
\begin{eqnarray}
  \frac{\delta \mathcal{L}(T_t q)}{\delta q(\tau)} &= & \dot q(t)\,\dot\delta(t-\tau)-\omega^{2}q(t)\,\delta(t-\tau) +\nonumber \\[1ex]  \label{e4x3a}
  & & \quad \frac{g}{4}\left[ q(t)\,k(t-\tau) + \delta(t-\tau)\!\int_{\mathbb{R}}\!d\sigma\,k(t-\sigma)q(\sigma)  \right]  \\[1ex]
  \frac{\delta^{2} \mathcal{L}(T_t q)}{\delta q(\tau)\,\delta q(\rho)} &= & \dot\delta(t-\rho)\,\dot\delta(t-\tau)
     -\omega^{2}\delta(t-\rho)\delta(t-\tau) + \nonumber \\[1ex]  \label{e4x3b}
  & &\quad \frac{g}{4}\,\left[k(t-\tau) \delta(t-\sigma) + k(t-\sigma) \delta(t-\tau) \right] \,,
\end{eqnarray}
and (\ref{nl19a}) yields
\begin{equation}\label{e4x3c}
     \Omega(q,\tau,\sigma) = - \ddot\delta(\tau-\sigma) - \omega^2\,\delta(\tau-\sigma) +\frac{g}2\,e^{-|\tau-\sigma|}   \,.
\end{equation}
Inserting (\ref{e4x3a}) into the EL equation~(\ref{nl4b}), we obtain
\begin{equation}\label{e4x4}
     \mathcal{E}(q,\tau) \equiv - \ddot q(\tau) - \omega^2 q(\tau) +\frac{g}2\!\int_\RR \D\sigma\,q(\sigma)\,e^{-|\tau-\sigma|} = 0\,.
\end{equation}
Similarly, substituting (\ref{e4x3c}) into the presymplectic form~(\ref{nl19b}) yields
\begin{equation}\label{e4x5a}
     \omega(q,\tau) = \delta P \wedge \delta Q + \delta \pi \wedge \delta \xi \,,
\end{equation}
where 
$Q,\,P,\,\xi\,$, and $\pi$ depend on the trajectory $q(\tau)\,$ through:
\begin{equation}\label{e4x5b}
\left. \begin{array}{ll}
    Q(q) := q(0)\,, \qquad & {\xi(q) := \sqrt{\frac{g}2}\,\int_0^{\infty} \D\sigma\,q(\sigma)\,e^{-\sigma}} \\[1.5ex]  
		P(q) := \dot q(0)\,, \qquad & {\pi(q) := \sqrt{\frac{g}2}\,\int_{-\infty}^0 \D\sigma\,q(\sigma)\,e^\sigma }		
    \end{array}   \right\}\,.
\end{equation}
Furthermore, using (\ref{e4x3a}) in expression~(\ref{nl20}), we find the energy expressed in terms of these canonical variables, thus obtaining the Hamiltonian:
\begin{equation}\label{e4x5c}
   H := \frac12\,\left(P^2 + \omega^2 Q^2 \right) - \sqrt{\frac{g}2}\,(\pi + \xi)\,Q + \pi \xi  	\,.
\end{equation}

Now, the 2-form \(\omega\) is defined on the infinite-dimensional space \(\mathcal{K}\) and is therefore degenerate. However, to determine whether the induced form \(j^*\omega\) is symplectic on the phase space \(\mathcal{D}\), we must examine the constraints that define this submanifold, namely the EL equation~(\ref{e4x4}). A straightforward calculation shows that 
\begin{equation}\label{e4x6}
 (\mathbf{D}^2 -1 ) \,\mathcal{E}(q,\tau) \equiv - \left[ (\mathbf{D}^2 -1 ) (\mathbf{D}^2 + \omega^2 ) + g \right] q(\tau) = 0 \,,
\end{equation}
for all $q\in \mathcal{D}\,$, which is a fourth order differential equation. Therefore,  $q$ depends on four parameters, at most, and $\dim\mathcal{D} \leq 4\,$ \footnote{Actually the reasoning is similar to that applied in Section \ref{PhSpEx}.}.

Additionally, the trajectory \( q(\tau) \) must satisfy the summability condition \( e^{-|\tau|} q(\tau) \in L^1(\mathbb{R}) \). This imposes further restrictions on the characteristic roots of equation~(\ref{e4x6}), specifically requiring that the absolute values of their real parts be strictly less than 1. A careful analysis shows that 
\begin{align*}
0 < 4g < (\omega^2 + 9) (\omega^2+1) & \quad |\realp(r)|< 1 \quad\mbox{for all characteristic roots and } \quad \dim\mathcal{D}= 4 \\[1.5ex] 
4g \geq (\omega^2 + 9) (\omega^2+1) & \quad |\realp(r)|\geq 1 \quad\mbox{for all characteristic roots and } \quad \dim\mathcal{D}= 0 \\[1.5ex] 
g \leq 0 & \quad \mbox{two characteristic roots fulfill } \quad  |\realp(r)|< 1\,, \quad \dim\mathcal{D}= 2\,. 
\end{align*}

Considering explicitly the four-dimensional case, when restricted to the phase space $\mathcal{D}$, the 2-form $\omega$ becomes symplectic (i.e., non-degenerate), and $Q,\,P,\,\xi\,$ and $\pi$ constitute a set of canonical coordinates. 
From the identity~(\ref{nl15y}), it immediately follows that, on the phase space, we have $\,i_\mathbf{D} \omega + \delta H = 0\,$, where \(\mathbf{D}\) denotes the restriction of the vector field to the submanifold \(\mathcal{D}\subset \mathcal{K}\), that is
$$ \mathbf{D} = \dot P\,\frac{\partial}{\partial P} + \dot Q\,\frac{\partial}{\partial Q} + \dot \pi\,\frac{\partial}{\partial \pi} +
 \dot \xi\,\frac{\partial}{\partial \xi} \,. $$
The Hamilton equations are then
\begin{eqnarray} \label{e4x7a}
  & & \hspace*{-8em} \dot Q =  P \,, \qquad \qquad \qquad \qquad \dot P = - \omega^{2}Q + \sqrt{\frac{g}{2}}\,(\pi+\xi) \\[2ex]\label{e4x7b}
  & & \hspace*{-8em} \dot\xi = \xi - \sqrt{\frac{g}{2}}\,Q  \,, \quad   \qquad\qquad \dot\pi = - \pi + \sqrt{\frac{g}{2}}\,Q \,,
\end{eqnarray}
 that on recombination yield
$$    Q^{(iv)}+(\omega^{2}-1)\,\ddot Q+(g-\omega^{2})\,Q=0 \,, $$
which coincides with (\ref{e4x6}). The general solution is thus
\begin{equation}\label{e4x8}
\left. \begin{array}{ll}
    Q(t) = \sum_{j=1}^4 A_j \,e^{r_j t}  \,, \qquad & P(t) = \sum_{j=1}^4 A_j\, r_j \,e^{r_j t}  \\[1.5ex]  
		\xi(t) = \sqrt{\frac{g}{2}}\,\sum_{j=1}^4 \frac{A_j}{1-r_j} \,e^{r_j t}\,, \qquad & \pi(t) = \sqrt{\frac{g}{2}}\,\sum_{j=1}^4 \frac{A_j}{r_j+1} \,e^{r_j t} \,
    \end{array}   \right\}\,.
\end{equation}

Substituting the explicit solution \( q(\tau) = \sum_{j=1}^{4} A_j\,e^{r_j \tau} \), which satisfies the Euler-Lagrange equation~(\ref{e4x4}), into the definitions~(\ref{e4x5b}), we obtain
\[
Q(T_t q) = Q(t), \quad P(T_t q) = P(t), \quad \xi(T_t q) = \xi(t), \quad \pi(T_t q) = \pi(t).
\]
This proves a 1-to-1 correspondence between solutions of the Euler-Lagrange equations and those of the Hamilton equations, and vice versa.

If $g\leq 0\,$, only two characteristic roots have the real part smaller than 1, namely,
$$ r_\pm = \pm i\,\omega_0\,, \qquad{\rm with}\qquad
\omega_0^2 = \sqrt{\frac{(1+\omega^2)^2}4+ |g|} + \frac{\omega^2-1}2  \,. $$
As a consequence, the phase space $\mathcal{D}$ is 2-dimensional and there are two relations connecting the
variables $Q,\,P,\,\xi,\,\pi\,$:
$$ \xi = i\,\sqrt{\frac{|g|}2}\,\frac{Q+P}{1+\omega_0^2} \,, \qquad
 \pi = i\,\sqrt{\frac{|g|}2}\,\frac{Q-P}{1+\omega_0^2} \,$$
that, substituted in (\ref{e4x5a}), yields
$$ \omega = M\,\delta P \wedge \delta Q  \,, \qquad {\rm with}\qquad M := 1 + \frac{|g|}{(1+\omega_0^2)^2} \,, $$
which is clearly symplectic.

\subsection{The Nonlocal Pais-Uhlenbeck Model \label{Exemple1}}
We now consider the fully nonlocal Pais-Uhlenbeck model \cite{PaisUhlenbeck1950}. This system is particularly compelling as it constitutes the simplest exactly solvable model that encapsulates all essential conceptual challenges inherent to higher-derivative theories (such as unbounded energies).  Specifically, we focus on the following nonlocal Lagrangian:
$$ \mathcal{L}= q\,F(D) q\,,\qquad{\rm  with} \qquad F(z) = \prod_{n=1}^\infty \left(1 + \frac{z^2}{\pi^2 n^2} \right) = \frac{\sinh z}{z}\,.
\, $$
Using Fourier integral methods, the operator \(F(D)\) can be recast into an explicit functional (nonlocal) form as shown in ref.~\cite{Heredia2024}:
\begin{equation} \label{ex1}
F(D) q(\tau) = K \ast q (\tau)\,, \qquad {\rm where} \qquad K(\sigma) = \frac12\,\theta\left(1 - |\sigma|\right) \,, 
\end{equation}
namely, 
$$ K \ast q (t) = \frac12\, \int_{-1+t}^{1+t} \D \tau \,q(\tau) \,.$$
The kinematic space consists of trajectories \( q \in \mathcal{C}^0(\mathbb{R}) \), and the functional derivatives of the Lagrangian are given by
\begin{equation} \label{ex2}
\lambda(q,\rho,\tau):=\frac{\delta \mathcal{L}\left(T_\rho q\right)}{\delta q(\tau)} = \delta(\rho-\tau)\,K\ast q(\rho) + K(\rho-\tau)\, q(\rho)
\end{equation}
and
\begin{equation} \label{ex2a}
 \frac{\delta^2 \mathcal{L}\left(T_\rho q\right)}{\delta q(\tau) \,\delta q(\sigma)} = \delta(\rho-\tau)\,K(\rho-\sigma) + K(\rho-\tau)\,\delta(\rho - \sigma) \,.
\end{equation}
Substituting~(\ref{ex2a}) into the definition~(\ref{nl19a}) yields
\begin{equation}  \label{ex2b}
 \Omega(q,\tau,\sigma) = \int_\RR \D \rho\,  \frac{\delta^2\mathcal{L}\left(T_\rho q\right)}{\delta q(\tau) \,\delta q(\sigma)} = 2\, K(\tau- \sigma) = \theta(1-|\tau-\sigma|) \,,
\end{equation}
where the kernel \( K \) given in~(\ref{ex1}) has been used. Furthermore, the EL equation is
\begin{equation}  \label{ex2z}
 \mathcal{E}(q,t) \equiv \int_{-1}^1 \D\tau\,q(t+\tau) = 0 \,,
\end{equation}
which is equivalent to its vanishing at $t=0$ and the vanishing of its derivative for all $t$, namely
\begin{equation}  \label{ex2y}
  \int_{-1}^1 \D\tau\,q(\tau) = 0 \qquad {\rm and} \qquad q(t+1) = q(t-1)  \,.
\end{equation}
Thus, the phase space $\mathcal{D}$ consists of periodic functions with period 2 and zero mean. Each trajectory is uniquely determined by specifying $q(\tau)\,, \;\, -1\leq \tau \leq 1\,$, with the additional conditions: (a)  $q(1)=q(-1)\,$, in order to warrant that $q$ is class $\mathcal{C}^0(\RR)\,$, and (b)  $\;\int_{-1}^1 \D\tau\,q(\tau) = 0 \,$.

Substituting equation~(\ref{ex2b}) into the expression for the presymplectic form~(\ref{nl19b}), we obtain:
$$ \omega = \int_0^\infty  \D \xi\, \int_{-\infty}^0 \D \tau\,\theta(1-|\xi-\tau|) \,\delta q(\tau) \wedge \delta q(\xi) \,. $$
Since $\xi \geq 0\,$, $\tau \leq 0\,$ and $|\xi-\tau|\leq 1\,$, the integration domain reduces to 
$\, \xi-1 \leq \tau \leq 0 \,$  and $\, 0 \leq \xi \leq 1 \,$. Therefore, the presymplectic form becomes
$$ \omega = \int_0^1  \D \xi\, \int_{\xi-1}^0 \D \tau\,\delta q(\tau) \wedge \delta q(\xi) \,, $$
which can be rewritten as
\begin{equation} \label{ex3}
\omega = \int_0^1  \D\xi\,\delta P(\xi)\wedge \delta Q(\xi) \,, 
\end{equation}
with $\;P(\xi):= \int_{\xi-1}^0 \D \tau\, q(\tau)\,$ and $\,Q(\xi) = q(\xi)\,, \;\, 0\leq \xi \leq 1\,$. 

In turn, the energy~(\ref{nl20}) takes the form \( E(q) = E_0(q) - \mathcal{L}(q) \), where
\begin{eqnarray}
 E_0(q) & := & \int_0^\infty \D\xi\, \int_{-\infty}^0 \D\tau\, \left\{ \left[\delta(\tau-\xi) \, K \ast q(\tau) + K(\tau-\xi) \,q(\tau)\right]\,\dot q(\xi)  - \right.   
\nonumber \\[1.5ex]
  & & \hspace*{5em} \left.   \left[\delta(\xi-\tau) \, K \ast q(\xi) + K(\xi-\tau) \,q(\xi)\right]\,\dot q(-\tau) \right\}  
	\nonumber \\[1.5ex]
  & \equiv & \frac12\,\int_0^1 \D\xi\, \int_{\xi-1}^0 \D\tau\, \left[q(\tau)\,\dot q(\xi) - q(\xi)\,\dot q(\tau) \right]	\,,
\end{eqnarray}
where the kernel \( K \) from~(\ref{ex1}) has been used, along with the fact that \( \tau \leq 0 \leq \xi \). In terms of the variables $Q$ and $P$, it yields
$$ E_0(q) \equiv \frac12\,\int_0^1 \D\xi\, \left[P(\xi)\,\dot Q(\xi) - Q(\xi)\,Q(0) - Q(\xi)\,\dot P(\xi) \right]  \,.$$
After an integration by parts, we get:
\begin{equation} \label{ex5}
E_0(q) \equiv \frac12\,\int_0^1 \D\xi\, \left[2 P(\xi)\,\dot Q(\xi) - Q(\xi)\,Q(0) \right] - \frac12\,Q(1) P(1) + \frac12\,Q(0) P(0)  \,.
\end{equation}

As for the Lagrangian, using equations~(\ref{ex1}) and~(\ref{ex2y}), we obtain:
\begin{equation} \label{ex5a}
\mathcal{L}(q) = q(0)\,\left(K \ast q\right)(0) = \frac12\,Q(0)\,\left[P(0) + \int_0^1 \D\tau\,Q(\tau)   \right] \,.
\end{equation}
Combining this with the expression for the energy, we arrive at the Hamiltonian:
\begin{equation} \label{ex5b}
H \equiv \int_0^1 \D\xi\, \left[P(\xi)\,\dot Q(\xi) -  Q(\xi)\, Q(0)\right] -\frac12\,Q(1)\,\Phi_1 \,.
\end{equation}

As a 2-form on $\mathcal{K}\,$,  $\,\omega $ is highly degenerate because it does not depend on the whole $q(\tau)\,$ but only on the piece $-1\leq \tau \leq 1\,$. However, the situation improves significantly when considering the restriction of $\omega$ to the space $\,(Q,P) \in \mathcal{C}^1([0,1],\RR^2)\,$, because~(\ref{ex3}) is the expression of a symplectic form in canonical coordinates.

As previously noted, each element $q \in \mathcal{D}\,$ is determined by its values on the interval $-1\leq \tau \leq 1\,$ , subject to only two additional constraints. Consequently,  the phase space can be embedded as a submanifold of $\mathcal{C}^1([0,1],\RR^2)\,$ via the map $\;j: \mathcal{D} \longrightarrow \mathcal{C}^1([0,1],\RR^2) \,$, defined by
\begin{equation}  \label{ex3z}
Q(\tau) = q(\tau)  \qquad {\rm and} \qquad P(\tau):= \int_{\tau-1}^0 \D \xi\, q(\xi) \,, \qquad 0\leq \tau \leq 1\,.
\end{equation}
Note that \( P(1) = 0 \) by construction. The inverse map recovers the trajectory \( q(\tau) \) from the pair \( (Q, P) \) as follows:
\begin{equation} \label{ex3b}
q(\tau) = \left\{\begin{array}{lcl}
                     Q(\tau) \,, &\quad & 0 \leq \tau \leq 1 \\
										 -\dot P(1+\tau) \,,& \quad & -1 \leq \tau \leq 0
										\end{array}    \right.  \,.
\end{equation}

Four constraints define the submanifold $\,j\left( \mathcal{D}\right) \,$, namely
\begin{eqnarray}    \label{ex3aa}
 \mathbf{1.-}\quad \Phi_1 \equiv P(1) = 0 \,, &\quad & \mbox{from (\ref{ex3z}) } \\[1.5ex]  \label{ex3a2}
 \mathbf{2.-}\quad \Phi_2 \equiv Q(0) + \dot P(1) = 0 \,, & & \mbox{$q$ is continuous at $\tau=0$} \\[1.5ex] \label{ex3a3}
 \mathbf{3.-}\quad \Phi_3 \equiv Q(1) + \dot P(0) = 0 \,, & & \mbox{$q$ is periodic and continuous} \\[1.5ex]  \label{ex3a4}
 \mathbf{4.-}\quad \Phi_4 \equiv P(0) + \int_0^1 \D\tau\,Q(\tau) = 0 \,, & & \mbox{EL equation} \,.
\end{eqnarray}

We have thus established a constrained Hamiltonian structure on \( \mathcal{C}^1([0,1], \mathbb{R}^2) \). As in the previous sections, the Hamilton equations follow from $\,i_\mathbf{D}\omega + \delta H = 0\,.$ However, due to the presence of constraints,  the 1-forms $\delta P(\xi)$ and $\delta Q(\rho)$ are not independent. To treat them as if they were, we introduce four Lagrange multipliers enforcing the constraints. The resulting Hamilton equations then take the form
$\,\displaystyle{ i_\mathbf{D}\omega + \delta H + \sum_{a=1}^4 \mu_a\,\delta\Phi_a = 0  }\,$,  on\; $\mathcal{D} \,$, or
\begin{eqnarray*}
 0 & = & \int_0^1 \D\xi \left\{\left[\mathbf{D}P(\xi) - \dot P(\xi) - Q(0)\right] \,\delta Q(\xi) + 
            \left[-\mathbf{D}Q(\xi) + \dot Q(\xi)\right] \,\delta P(\xi) \right\} + \\[1.5ex]
  & & + \left[\mu_1 - \frac12 Q(1) \right]\,\delta \Phi_1 + \sum_{a=2}^4 \mu_a\,\delta\Phi_a \,,
\end{eqnarray*}
which determines the Lagrange multipliers: 
$$\mu_1 = \frac12\,Q(1)\, \qquad {\rm and} \qquad \mu_a = 0 \,, \;\, a=2,3,4\,,$$ 
and yields the equations:
\begin{equation} \label{ex7a}
 \mathbf{D}P(\xi) - \dot P(\xi) - Q(0) = 0 \qquad {\rm and} \qquad \mathbf{D}Q(\xi) -\dot Q(\xi) = 0 \qquad\mbox{for $0 < \xi < 1\,$,}
\end{equation}
which form a differential system on the fields $\,\left(Q(\xi),P(\xi)\right)\,$. The solution is a curve on the phase space
$$ t\in \RR \longrightarrow \left( P(t,\xi), Q(t,\xi)\,\right) \in j(\mathcal{D}) $$
satisfying
\begin{equation} \label{ex7c}
 \partial_t Q(t,\xi) - \partial_\xi Q(t,\xi) = 0 \,, \qquad \quad \partial_t P(t,\xi) - \partial_\xi P(t,\xi) - Q(t,0) = 0 \,,
\end{equation} 
and fulfilling the constraints
\begin{equation} \label{ex7z}
\left.
\begin{array}{lcl}
\Phi_1(t) \equiv P(t,1) = 0 \,, & \qquad & \Phi_2(t) \equiv Q(t,0) + \dot P(t,1) = 0  \,, \\[1.5ex]
\Phi_3(t) \equiv Q(t,1) + \dot P(t,0) = 0  \,, & \qquad & \Phi_4(t) \equiv P(t,0) + \int_0^1 \D\tau \,Q(t,\tau) = 0  \,
\end{array}  \right\}  \,,
\end{equation}
which means that the curve remains on the submanifold defined by the constraints.

The general solution to the system of partial differential equations~(\ref{ex7c}) is
$$ Q(t,\xi) = f(t+\xi) \qquad {\rm and} \qquad  P(t,\xi) = g(t+\xi) + \int_0^t\D\tau\,f(\tau) \,, \qquad 0 \leq \xi \leq 1 \,,\quad t\in \RR \,,$$  
where $f(\tau)$ and $g(\tau)$ are two arbitrary functions in $\mathcal{C}^1([0,1],\RR)\,$. Now, imposing the constraint \( \Phi_1(t) = 0 \), we obtain
$$ g(\xi) = \int_{\xi-1}^0 \D\tau\,f(\tau) \,, $$
so that the general solution becomes:
\begin{equation} \label{ex7d}
 Q(t,\xi) = f(t+\xi) \qquad {\rm and} \qquad  P(t,\xi) =  \int_{\xi-1}^0\D\tau\,f(t+\tau) \,. 
\end{equation}
Substituting this expression into \( \Phi_2(t) = 0 \) does not impose any further restriction on \( f(t) \).

Next, substituting~(\ref{ex7d}) into \( \Phi_3(t) = 0 \) and \( \Phi_4(t) = 0 \) yields, respectively,
$$ f(t+1) = f(t-1) \qquad {\rm and} \qquad \int_{-1}^1 \D\tau\, f(t+\tau) = 0\,,$$
which are equivalent to the time-independent conditions:
$$ f(t+1) = f(t-1) \qquad {\rm and} \qquad \int_{-1}^1 \D\tau\, f(\tau) = 0 \,.$$
Thus $f(\tau)\,$ is a solution of (\ref{ex2y}) and the solutions of the Hamilton equations and the EL equation are in 1-to-1 correspondence.

\subsection{A ``delayed'' Harmonic Oscillator \label{Exemple2}}
We introduce a minimal nonlocal deformation of the ordinary harmonic oscillator to emulate how a fixed temporal separation can describe (memory) effects. Specifically, we consider
\begin{equation} \label{e2x1}
\mathcal{L}\left(T_t q\right) = \frac12\,\dot q^2(t) -\frac12\, q^2(t) + \kappa\,q(t)\,q(t+T) \,,
\end{equation}
where the bilocal term $q(t)\,q(t+T)$ correlates the present state with its value a later time, $t+T$, providing a feedback mechanism. Its functional derivatives are
\begin{equation} \label{e2x2}
\lambda(q,\rho,\tau):= \dot q(\rho)\,\dot\delta(\rho-\tau) - q(\rho)\,\delta(\rho-\tau) + \kappa\,\left[q(\rho+T)\,\delta(\rho-\tau) + q(\rho)\,\delta(\rho+T-\tau) \right] \,
\end{equation}
and
\begin{eqnarray} 
 \frac{\delta^2 L\left(T_\rho q\right)}{\delta q(\tau) \,\delta q(\sigma)} & = & \dot\delta(\rho-\tau)\,\dot\delta(\rho-\sigma) - \delta(\rho-\tau)\,\delta(\rho-\sigma) +\kappa\,\left[\delta(\rho+T-\sigma)\,\delta(\rho-\tau) + \right.   \nonumber\\[1.5ex] \label{e2x2a}
  & & \left. \delta(\rho-\sigma)\,\delta(\rho+T-\tau) \right] \,
\end{eqnarray}
that, substituted in (\ref{nl19a}), yields
\begin{equation}  \label{e2x2b}
 \Omega(q,\tau,\sigma) = -\ddot\delta(\sigma-\tau) - \delta(\sigma-\tau) + \kappa\,\left[\delta(\tau+T-\sigma) + \delta(\sigma+T-\tau) \right]\,.
\end{equation}
The EL equation (\ref{nl4}) is
\begin{equation}  \label{e2x2z}
 \mathcal{E}(q,t) \equiv - \ddot q(t) - q(t) + \kappa \,\left[q(t+T)+ q(t-T) \right]  = 0 \,. 
\end{equation}
Notice that the presence of second derivatives and time-shifted arguments immediately implies that any continuous solution $q(t)$ satisfying (\ref{e2x2z}) must necessarily be smooth, as higher-order derivatives can be recursively determined from the equation itself.

This equation allows to determine the value of $q(t+T)$ from the piece of trajectory in the past of $t+T$ (respectively, $q(t-T)$ in terms of the piece of trajectory in the future of $t-T$).
As a consequence, an iterative procedure can be implemented to determine $q(\tau+nT)$, with $0\leq \tau\leq T$, explicitly in terms of the previous values $q(\tau+[n-1]T)$ and $q(\tau+[n-2]T)$ for integers $n\geq 1$ (respectively, $q(\tau+nT)$ in terms of subsequent values $q(\tau+[n+1]T)$ and $q(\tau+[n+2]T)$ for integers $n\leq -2$). Thus, a solution to the EL equation is fully determined by specifying the function $q(\tau)$ on an interval of length $2T$, for example $|\tau|<T$. However, ensuring continuity of $q(t)$ for all times imposes additional constraints, linking values of $q$ and its derivatives at the endpoints $\pm T$. These constraints arise naturally by evaluating equation (\ref{e2x2z}) and its time derivatives at $t=0$.

Substituting (\ref{e2x2b}) into the symplectic form (\ref{nl19b}), we arrive, after some algebra, at
\begin{equation} \label{e2x3}
\omega = \delta \pi \wedge \delta \xi + \int_0^T  \D\tau\,\delta P(\tau)\wedge \delta Q(\tau) \,, 
\end{equation}
where we have introduced  $\,\pi := \dot q(0)\,, \,\, \xi := q(0)\,$, and defined new canonical fields by
\begin{equation} \label{e2x3a}
\,Q(\tau) := \sqrt{\kappa} \,q(\tau)\qquad {\rm and} \qquad P(\tau) := \sqrt{\kappa} \,q(\tau-T)\,, \qquad  0\leq \tau \leq T\,.
\end{equation}
We have assumed here $\kappa > 0$. If instead $\kappa<0$, one must replace $\sqrt{\kappa}$ with $\sqrt{|\kappa|}$ and exchange the roles of $Q$ and $P$.

In the expression for $\omega$ given above, the symplectic structure is manifest, and the associated set of canonical variables ---as well as the elementary Poisson brackets--- can be immediately read off. The definitions in (\ref{e2x3a}) identify $Q(\tau)$ and $P(\tau)$ as field variables constructed from the segment of the trajectory $q(\tau)$ restricted to the interval $-T \leq \tau \leq T$:
\begin{equation} \label{e2x3b}
q(\tau) = \left\{\begin{array}{lcl}
                     \kappa^{-1/2} Q(\tau) \,, &\quad & 0 \leq \tau \leq T \\[1.5ex] 
										 \kappa^{-1/2} P(T+\tau) \,,& \quad & -T \leq \tau \leq 0
										\end{array}    \right.  \,.
\end{equation}

Five constraints on the canonical variables follow from the definition (\ref{e2x3a}), the continuity condition at $\tau=0$ in (\ref{e2x3b}), and the EL equation (\ref{e2x2z}), namely
\begin{equation}\label{e2x4}
\left.
\begin{array}{lcl}
\phi_1 &:=& \sqrt{\kappa}\,\xi - Q(0) = 0, \\[1.5ex]
\phi_2 &:=& \sqrt{\kappa}\,\xi - P(T) = 0, \\[1.5ex]
\phi_3 &:=& \sqrt{\kappa}\,\pi - \dot Q(0) = 0, \\[1.5ex]
\phi_4 &:=& \sqrt{\kappa}\,\pi - \dot P(T) = 0, \\[1.5ex]
\phi_5 &:=& -\ddot Q(0) - Q(0) + \kappa\left[Q(T) + P(0)\right] = 0.
\end{array}
\right\}
\end{equation}

The Hamiltonian, expressed in terms of the canonical variables, reads:
\begin{equation} \label{e2x5}
H \equiv \frac12\,\pi^2 + \frac12\,\xi^2 +\int_0^T \D\tau\, P(\tau)\,\dot Q(\tau) -  P(T)\, Q(T) \,. 
\end{equation}
The phase space is thus coordinatized by the pair of real numbers $\pi$ and $\xi$, together with the fields $Q(\tau)$ and $P(\tau)$ of class $\mathcal{C}^\infty$, with $0 \leq \tau \leq T\,$, all subject to the constraints (\ref{e2x4}). The time evolution of the system is governed by the Hamiltonian (\ref{e2x5}).

As in the previous example, the Hamilton equations must include one Lagrange multiplier for each constraint. Thus, the dynamics is governed by 
\begin{equation} \label{e2x6}
 i_\mathbf{D}\omega + \delta H + \sum_{a=1}^5 \lambda_a\,\delta \phi_a = 0  \,,  
\end{equation}
where summation over repeated indices is understood. After a straightforward computation, one finds that
\begin{equation} \label{e2x6a}  
i_\mathbf{D}\omega + \delta H +  \sum_{a=1}^5 \lambda_a\,\delta \phi_a \equiv  
   a\,\delta\xi + b\,\delta \pi + \int_0^T \D\tau\,\left[ A(\tau) \,\delta Q(\tau) + B(\tau) \,\delta P(\tau) \right] \,, 
\end{equation}
where the coefficients are given by:
\begin{eqnarray*}  
a & := & \displaystyle{\mathbf{D}\pi + \xi +\kappa^{1/2} \left[\lambda_1+\lambda_2\right]} \\[1.5ex]
b & := & \displaystyle{-\mathbf{D}\xi + \pi +\kappa^{1/2} \left[\lambda_3+\lambda_4\right]} \\[1.5ex]
 A(\tau) & := & \displaystyle{\mathbf{D}P(\tau) - \dot P(\tau) - \left[\lambda_1 + \lambda_5 + P(0)\right]\,\delta(\tau) +\lambda_3\,\dot\delta(\tau) -\lambda_5\ddot\delta(\tau) + \kappa\,\lambda_5\;\delta(\tau-T) } \\[1.5ex]
 B(\tau) & := & \displaystyle{-\mathbf{D}Q(\tau) +\dot Q(\tau) - (Q(T) + \lambda_2)\,\delta(\tau-T) +\lambda_4\,\dot\delta(\tau-T) + \lambda_5\,\kappa\,\delta(\tau)} \,.
\end{eqnarray*}
Thus, equation (\ref{e2x6a}) implies the vanishing of $a$, $b$, $A(\tau)$, and $B(\tau)$ on the constraint submanifold. If the endpoints of the interval are temporarily excluded, this leads to
\begin{equation} \label{e2x7a}
 \mathbf{D}P(\tau) - \dot P(\tau) = 0 \qquad {\rm and} \qquad \mathbf{D}Q(\tau) -\dot Q(\tau) = 0 \,, \qquad 0 < \tau < T\,.
\end{equation}
However, since $P(\tau)$ and $Q(\tau)$ are continuous at the endpoints, these equations also hold at $\tau=0$ and $\tau=T$, so that they are valid throughout the entire interval $[0,T]$. The vanishing of the generalized functions $A(\tau)$ and $B(\tau)$ then further implies
\begin{equation} \label{e2x7y}
 \lambda_1= - P(0)\,, \qquad \lambda_2= -Q(T)\,,   \qquad {\rm and} \qquad \lambda_3 = \lambda_4 = \lambda_5 = 0 \, .
\end{equation}
Substituting these values into the conditions $a = 0$ and $b = 0$ yields
$$  \mathbf{D}\pi +\xi - \kappa^{1/2} \left[P(0)+Q(T) \right]  = 0 \qquad {\rm and} \qquad 
\mathbf{D}\xi -\pi = 0 \,,$$
which together amount to
\begin{equation} \label{e2x7b}
\pi = \mathbf{D}\xi \qquad {\rm and} \qquad 
 \mathbf{D}^2\xi +\xi - \kappa^{1/2} \left[P(0)+Q(T) \right]  = 0  \,.
\end{equation}

The solution to equations (\ref{e2x7a}) and (\ref{e2x7b}) consists of some functions $\xi(t)$, $\pi(t)$, $Q(t,\tau)$, and $P(t,\tau)$ satisfying the constraints (\ref{e2x4}), namely
$$  Q(t,0) = P(t,T) =\sqrt{\kappa} \,\xi(t) = 0  \qquad {\rm and} \qquad 
\dot Q(t,0) = \dot P(t,T) = \sqrt{\kappa}\,\pi(t) = 0\,. $$
The general solution to (\ref{e2x7a}) is given by
\begin{equation} \label{e2x7}
 P(t,\tau) = g(t+\tau) \qquad {\rm and} \qquad Q(t,\tau) = f(t+\tau) \,, \qquad t \in \RR \,, \quad 0 < \tau < T \,, 
\end{equation}
where $f$ and $g$ are arbitrary smooth functions of a single real variable. Imposing the constraints, one obtains
$$ \xi(t) = \kappa^{-1/2} f(t) \qquad {\rm and } \qquad f(t) = g(t+T) \,, $$
which, substituted into equation (\ref{e2x7b}), gives the nonlocal evolution equation
\begin{equation} \label{e2x7z}
 \ddot f(t)  + f(t) - \kappa  \left[f(t-T) + f(t+T) \right]  = 0  \,,
\end{equation}
namely, the EL equation (\ref{e2x2z}).

\section{Conclusion}
The main goal of this work has been to construct a Hamiltonian formalism for a nonlocal Lagrangian system. This requires: (1) establishing a phase space, (2) finding the Hamiltonian and the canonical momenta, and (3) defining the Poisson brackets of any pair of functions defined on this phase space ---or, alternatively, defining a symplectic structure (dual of the PB) on it. For this last step, it is essential to properly define the Legendre transformation to identify the canonical momenta.

For guidance, and in order to imitate it ---mutatis mutandis--- in the nonlocal case, we have reviewed how the Hamiltonian formalism is constructed in the local cases: for Lagrangians that depend on coordinates and velocities ---the standard undergraduate-level case--- and also for Lagrangians that involve derivatives up to the $n$-th order. For a regular first-order Lagrangian with $m$ degrees of freedom, the Euler-Lagrange equations form a second-order ordinary differential system that can be written in normal form; then the existence and uniqueness theorems establish that each solution is determined uniquely by $2 m$ initial conditions. Thus, the Euler-Lagrange equations can be regarded as evolution equations in the state space (or phase space), starting from an initial state. 

The previous interpretation is the one commonly presented in textbooks. However, an alternative viewpoint is also possible: the Euler-Lagrange equations can be understood as constraints delimiting the subset of dynamic trajectories of the system within the class of all kinematically possible trajectories. 

In the local case, the canonical momenta can be defined in two equivalent ways: (a) as the boundary terms arising from integration by parts in the variational principle, and (b) as ingredients in the conserved quantities resulting from Noether's theorem\footnote{Both approaches are closely related, since Noether's theorem itself relies on a variation of the action integral, although these variations are different in nature from those that lead to the Euler-Lagrange equations.}. 

We then addressed the case of nonlocal Lagrangian systems by using the functional formalism, which primarily involves the use of integral operators. In such systems, the Lagrangian depends explicitly on the entire trajectory. If the action integral is over a finite time interval $\,[t_1,t_2]\,$, the action will also depend on the entire trajectory, and not just on the part corresponding to the integration interval\footnote{Something similar happens if the Lagrangian depends on a bounded piece of the trajectory and not just on an instant.}. Therefore, we first had to reformulate the variational principle for nonlocal Lagrangians, in a way that avoids this issue and the associated inconsistencies discussed in Section \ref{Paradox}. 

In the nonlocal case, the Euler-Lagrange equations are not an ordinary differential system; rather, they form a system of implicit functional equations ---often of integro-differential nature--- for which no general theorems on the existence and uniqueness of solutions are available. In this scenario, only the second view mentioned in the local case is viable: the Euler-Lagrange equations are implicit equations (constraints) defining the submanifold of dynamic trajectories $\mathcal{D}$ ---the phase space--- as a subset of all kinematic trajectories, $\mathcal{K}\,$.

In general, the derivation of the Euler-Lagrange equation in the nonlocal case does not involve any integration by parts, which is typically used to recognize the canonical momenta. Therefore, to obtain them, we must resort to a generalization of Noether's theorem for nonlocal Lagrangians, which we have proven here. Specifcally, the form of the conserved quantity associated with an infinitesimal symmetry transformation provides us with the information to infer the definition of the canonical momenta. 

This approach also provides us with the definition of energy as the conserved quantity associated with the time-translation symmetry. Such a definition remains valid regardless of whether or not there is invariance; the only consequence of this fact is that energy may not be conserved.

Our approach to the Hamiltonian formalism is based on the equation $\,i_\mathbf{D} \omega + \delta E = 0\,$, where $\mathbf{D}$ is the generator of the time evolution, $E$ is the energy and $\omega $ is the presymplectic form, a 2-differential form that is constructed from the canonical momenta. We have derived explicit expressions for both $\omega$ and $E$ on the kinematic space $\mathcal{K}$, given by equations (\ref{nl19b}) and (\ref{nl20}), respectively.

The crucial point, which culminates the whole process, consists of restricting the form $\omega$ to the phase space $\mathcal{D}$ and subsequently showing that this restricted form is symplectic (i.e., non-degenerate)\footnote{In the worst case, if this restriction is degenerate, the result would be a Hamiltonian formalism with constraints.}. Once this has been established, one proceeds by finding a canonical coordinate system in the phase space. Finally, by expressing the restriction of $E$ in these coordinates, the Hamiltonian is obtained explicitly. 

In the case of a local Lagrangian ---considered here as a particular case---, our method yields the same result as the usual standard procedure. For the rest, however, we do not yet have a general systematic way to show that the resulting 2-form is symplectic. Nevertheless, we have successfully applied our method to three illustrative examples, discussed in Section~\ref{S6}, which provide practical guidance for tackling more general cases. The key step lies in how the Euler-Lagrange equations are implemented to characterize the phase space as a submanifold of $\mathcal{K}\,.$ Importantly, as demonstrated by these examples, it is not necessary to explicitly solve the Euler-Lagrange equations to verify that the 2-form is symplectic ---a significant improvement over previous approaches~\cite{Heredia2021a}. Finally, this method extends naturally to field theory by following the approach outlined in Ref.~\cite{Heredia2022}.

\section*{Acknowledgement}
J. Ll. was supported by the Spanish MINCIU and ERDF (project ref. PID2021-123879OBC22). C. H. thanks the IAMM group for their ongoing support.

\printbibliography

@article{Llosa1987,
  author    = {Xavier Ja{\'e}n and Rafael J{\'a}uregui and Josep Llosa and Antoni Molina},
  title     = {Hamiltonian formalism for path-dependent Lagrangians},
  journal   = {Physical Review D},
  volume    = {36},
  number    = {10},
  pages     = {2385--2398},
  month     = oct,
  year      = {1987},
}

@article{Vives1994,
  author    = {Josep Llosa and Jordi Vives},
  title     = {Hamiltonian formalism for nonlocal Lagrangians},
  journal   = {Journal of Mathematical Physics},
  volume    = {35},
  number    = {6},
  pages     = {2856--2877},
  month     = jun,
  year      = {1994},
}

@article{Gomis2002,
  author    = {Joaquim Gomis and Kiyoshi Kamimura and Josep Llosa},
  title     = {Hamiltonian formalism for space--time noncommutative theories},
  journal   = {Physical Review D},
  volume    = {63},
  number    = {4},
  pages     = {045003},
  month     = jan,
  year      = {2001},
}

@article{Heredia2021a,
  author  = {Carlos Heredia and Josep Llosa},
  title   = {Non-local Lagrangian mechanics: Noether’s theorem and Hamiltonian formalism},
  journal = {J. Phys. A: Math. Theor.},
  volume  = {54},
  pages   = {425202},
  year    = {2021},
}

@article{Heredia2022,
  author  = {Carlos Heredia and Josep Llosa},
  title   = {Nonlocal Lagrangian fields: Noether’s theorem and Hamiltonian formalism},
  journal = {Phys. Rev. D},
  volume  = {105},
  pages   = {126002},
  year    = {2022},
  month   = jun,
}

@book{Goldstein,
  author    = {Herbert Goldstein},
  title     = {Classical Mechanics},
  edition   = {2nd},
  publisher = {Addison‐Wesley},
  address   = {Reading, MA},
  year      = {1980}
}

@book{Gantmacher,
  author    = {F.~R. Gantmacher},
  title     = {Lectures in Analytical Mechanics},
  publisher = {Mir Publishers},
  address   = {Moscow},
  year      = {1975},
}

@book{Godbillon,
  author    = {Claude Godbillon},
  title     = {Géométrie différentielle et mécanique analytique},
  publisher = {Hermann},
  address   = {Paris},
  year      = {1969}
}

@book{Marsden,
  author    = {Jerrold E. Marsden and Tudor S. Ratiu},
  title     = {Introduction to Mechanics and Symmetry},
  publisher = {Springer},
  address   = {New York},
  year      = {1994}
}

@book{Barbu2016,
  author    = {Viorel Barbu},
  title     = {Differential Equations: An Introduction to Basic Concepts, Results and Applications},
  publisher = {Springer},
  address   = {Cham},
  year      = {2016},
  % isbn    = {978-3-319-45465-0},
  % doi     = {10.1007/978-3-319-45466-7},
}

@book{Arnold1992,
  author    = {V. I. Arnold},
  title     = {Ordinary Differential Equations},
  edition   = {2nd},
  publisher = {Springer},
  address   = {Berlin},
  year      = {1992},
  % note    = {(Traducción al inglés revisada)},
  % doi     = {10.1007/978-3-662-00942-7},
}

@article{Noether,
  author    = {Emmy Noether},
  title     = {Invariante Variationsprobleme},
  journal   = {Nachrichten von der Gesellschaft der Wissenschaften zu Göttingen, Mathematisch-Physikalische Klasse},
  pages     = {235--257},
  year      = {1918}
}

@article{Ostrogradski,
  author    = {M.~Ostrogradski},
  title     = {Mémoire sur les équations différentielles relatives au problème des isopérimètres},
  journal   = {Mem. Acad. St. Petersbourg},
  volume    = {6},
  pages     = {385--517},
  year      = {1850}
}

@book{Whittaker,
  author    = {Edmund T. Whittaker},
  title     = {A Treatise on the Analytical Dynamics of Particles and Rigid Bodies},
  edition   = {4th},
  publisher = {Cambridge University Press},
  year      = {1937}
}

@article{Fokker,
  author    = {Adriaan D. Fokker},
  title     = {Über ein Relativitätsprinzip der Elektrodynamik},
  journal   = {Zeitschrift für Physik},
  volume    = {58},
  pages     = {386--393},
  year      = {1929}
}

@article{WheelerFeynman,
  author    = {John A. Wheeler and Richard P. Feynman},
  title     = {Interaction with the Absorber as the Mechanism of Radiation},
  journal   = {Reviews of Modern Physics},
  volume    = {17},
  pages     = {157--181},
  year      = {1945}
}

@article{PaisUhlenbeck1950,
  author    = {Abraham Pais and George E. Uhlenbeck},
  title     = {On Field Theories with Non‐Localized Action},
  journal   = {Physical Review},
  volume    = {79},
  pages     = {145--165},
  year      = {1950}
}

@article{Yukawa,
  author    = {Hideki Yukawa},
  title     = {Quantum Theory of Non-Local Fields. Part I. Free Fields},
  journal   = {Physical Review},
  volume    = {77},
  pages     = {219--226},
  year      = {1950}
}

@article{Moller,
  author    = {P.~Kristensen and C.~Møller},
  title     = {On the Theory of the Electromagnetic Field in the Presence of Point Charges},
  journal   = {K. Dan. Vidensk. Selsk. Mat.-Fys. Medd.},
  volume    = {27},
  pages     = {7},
  year      = {1952}
}

@article{EliezerWoodard1989,
  author    = {David A. Eliezer and R.~P. Woodard},
  title     = {The Problem of Nonlocality in String Theory},
  journal   = {Nuclear Physics B},
  volume    = {325},
  pages     = {389--469},
  year      = {1989}
}

@article{Marnelius1973,
  author    = {Marnelius, R.},
  title     = {Action principle and nonlocal field theories},
  journal   = {Physical Review D},
  volume    = {8},
  number    = {8},
  pages     = {2472--2483},
  month     = oct,
  year      = {1973},
}

@book{Borel,
  author    = {Émile Borel},
  title     = {Leçons sur les Séries Divergentes},
  publisher = {Gauthier-Villars},
  address   = {Paris},
  year      = {1901}
}

@book{Apostol,
  author    = {Tom M. Apostol},
  title     = {Mathematical Analysis},
  publisher = {Addison‐Wesley},
  year      = {1957}
}

@article{Heydeman2023,
  author  = {Matthew Heydeman and Christian B. Jepsen and Ziming Ji and Amos Yarom},
  title   = {Polyakov’s confinement mechanism for generalized Maxwell theory},
  journal = {J.High Energy Phys.},
  volume  = {04},
  year    = {2023},
  pages   = {119},
}

@article{Choquet,
  author    = {Gustave Choquet},
  title     = {Theory of Capacities},
  journal   = {Annales de l'Institute Fourier},
  volume    = {5},
  pages     = {131--295},
  year      = {1954}
}

@article{JLlosaMT,
  author  = {Carlos Heredia and Josep Llosa},
  title   = {Energy-momentum tensor for the electromagnetic field in a dispersive medium as an application of Noether theorem},
  journal = {J. Phys. Commun.},
  volume  = {5},
  number  = {5},
  pages   = {055003},
  year    = {2021},
}

@misc{JLlosaMT-ArXiv,
  author       = {Carlos Heredia and Josep Llosa},
  title        = {Energy-momentum tensor for the electromagnetic field in a dispersive medium as an application of Noether theorem},
  howpublished = {arXiv:2002.12725v1},
  year         = {2020},
  note         = {Versión 1: 14 Feb 2020},
  url          = {https://arxiv.org/abs/2002.12725v1},
}

@book{Landau,
  author    = {Lev D. Landau and Evgeny M. Lifshitz},
  title     = {Mechanics},
  series    = {Course of Theoretical Physics},
  volume    = {1},
  publisher = {Pergamon Press},
  year      = {1976}
}

@article{Barnaby,
  author  = {Barnaby, Neil},
  title   = {A new formulation of the initial value problem for nonlocal theories},
  journal = {Nuclear Physics B},
  volume  = {845},
  number  = {1},
  pages   = {1--29},
  month   = apr,
  year    = {2011},
}

@book{Vladimirov,
  author    = {Vladimir S. Vladimirov},
  title     = {Equations of Mathematical Physics},
  publisher = {MIR Publishers},
  year      = {1984},
  note      = {Section 8.4, pp. 139}
}

@book{Flanders,
  author    = {Harley Flanders},
  title     = {Differential Forms with Applications to the Physical Sciences},
  publisher = {Dover},
  year      = {1989}
}

@article{Heredia2024,
  author    = {Carlos Heredia and Josep Llosa},
  title     = {Are nonlocal Lagrangian systems fatally unstable?},
  journal   = {arXiv preprint},
  volume    = {arXiv:2403.19777},
  year      = {2024},
}

@article{Witten1986,
  author  = {Witten, E.},
  title   = {Non‐commutative geometry and string field theory},
  journal = {Nuclear Physics B},
  volume  = {268},
  number  = {2},
  pages   = {253--294},
  year    = {1986},
}

@article{Biswas2012,
  author  = {Biswas, Tirthabir and Gerwick, Erik and Koivisto, Tomi and Mazumdar, Anupam},
  title   = {Towards singularity- and ghost-free theories of gravity},
  journal = {Physical Review Letters},
  volume  = {108},
  number  = {3},
  pages   = {031101},
  year    = {2012},
  month   = jan,
}

@article{HerediaKolar2022,
  author  = {Heredia, Carlos and Kolář, Ivan and Llosa, Josep and Maldonado Torralba, Francisco J. and Mazumdar, Anupam},
  title   = {Infinite-derivative linearized gravity in convolutional form},
  journal = {Classical and Quantum Gravity},
  volume  = {39},
  pages   = {085001},
  year    = {2022},
}

@article{BoosKolar2021,
  author  = {Boos, Jens and Kolář, Ivan},
  title   = {Nonlocality and gravitoelectromagnetic duality},
  journal = {Physical Review D},
  volume  = {104},
  number  = {2},
  pages   = {024018},
  year    = {2021},
  month   = jul,
}

@article{CalcagniModesto2014,
  author  = {Calcagni, Gianluca and Modesto, Leonardo},
  title   = {Nonlocality in string theory},
  journal = {Journal of Physics A: Mathematical and Theoretical},
  volume  = {47},
  number  = {35},
  pages   = {355402},
  year    = {2014},
  month   = aug,
}

@article{CapozzielloBajardi2022,
  author  = {Capozziello, Salvatore and Bajardi, Francesco},
  title   = {Nonlocal gravity cosmology: An overview},
  journal = {International Journal of Modern Physics D},
  volume  = {31},
  number  = {06},
  pages   = {2230009},
  year    = {2022},
}

@book{Mashhoon2017,
  author    = {Mashhoon, Bahram},
  title     = {Nonlocal Gravity},
  publisher = {Oxford University Press},
  address   = {Oxford, UK},
  year      = {2017},
  month     = jul,
}

@article{Tetrode1922,
  author    = {Tetrode, H.},
  title     = {Zur Quantelung der Wellenfelder},
  journal   = {Zeitschrift für Physik},
  volume    = {10},
  number    = {1},
  pages     = {317--338},
  year      = {1922},
  note      = {in German},
}

@book{CartanH,
  author    = {Henri Cartan},
  title     = {Formes différentielles},
  publisher = {Hermann},
  address   = {Paris},
  year      = {2007}
}

\end{document}